\documentclass[final,authoryear,3p,times,twocolumn]{elsarticle}

\usepackage{amssymb}

\biboptions{sort&compress}

\journal{Applied Radiation and Isotopes}

\begin{document}

\begin{frontmatter}



\title{Investigation of activation cross-sections of proton induced nuclear reactions on $^{nat}Tl$ up to 42 MeV: review, new data and evaluation}


\author[1]{F. T\'ark\'anyi}
\author[1]{F. Ditr\'oi\corref{*}}
\author[2]{A. Hermanne}
\author[1]{S. Tak\'acs}

\author[2]{R. Adam-Rebeles} 
\author[2]{N. Walravens}
\author[2]{O. Cichelli} 
\author[3]{A.V.  Ignatyuk}
\cortext[*]{Corresponding author: ditroi@atomki.hu}

\address[1]{Institute of Nuclear Research of the Hungarian Academy of Sciences (ATOMKI),  Debrecen, Hungary}
\address[2]{Cyclotron Laboratory, Vrije Universiteit Brussel (VUB), Brussels, Belgium}
\address[3]{Institute of Physics and Power Engineering (IPPE), Obninsk, Russia}
\begin{abstract}
Cross-sections of proton induced nuclear reactions on natural thallium have been studied for investigation of the production of the medical important $^{201}Tl$ diagnostic radioisotope. The excitation functions of $^{204m}Pb$, $^{203}Pb$, $^{202m}Pb$, $^{201}Pb$, $^{200}Pb$, $^{199}Pb$, $^{202}Tl$ (direct, cumulative), $^{201}Tl$ (direct, cumulative), $^{200}Tl$(direct), and $^{203}Hg$ were measured up to 42 MeV proton energy by stacked foil technique and activation method. The experimental data were compared with the critically analyzed experimental data in the literature, with the IAEA recommended data and with the results of model calculations by using the ALICE-IPPE, EMPIRE-II and TALYS codes.
\end{abstract}


\end{frontmatter}


\section{Introduction}
\label{1}

The radionuclide $^{201}Tl$ ($T_{1/2} = 73 h$) as a good potassium analog has been used widely in diagnostic nuclear medicine for forty years \citep{51}. $^{201}Tl$ decays by EC (100\%) and emits a 167 keV  $\gamma$-ray used for SPECT imaging. The $^{201}Tl$ can be produced at charged particle accelerators. The investigated nuclear reactions are collected in Table. 1 \citep{2,4,6,7,8,14,17,20,23,29,37,40,41,42,43,47,52,54,64}. The theoretical excitation functions for the most important routes are shown in Fig. 1. 

\begin{figure}[h]
\includegraphics[scale=0.3]{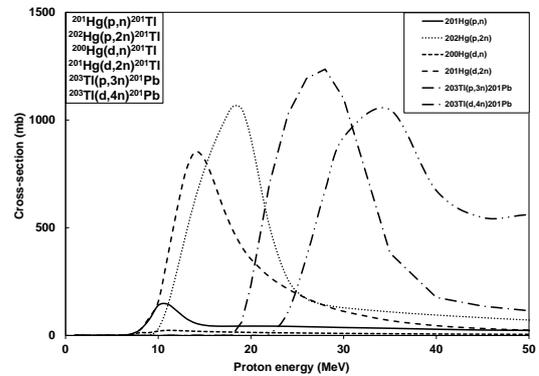}
\caption{Excitation functions (TENDL 2011) of the most important nuclear reactions for production of $^{201}Tl$}
\end{figure}

\begin{table*}[t]
\tiny
\caption{Investigated nuclear reactions for production of ${}^{201}$Tl }
\centering
\begin{center}
\begin{tabular}{|p{0.9in}|p{2.1in}|p{1.5in}|} 
\hline 
Reaction & References & Comment \\ \hline 
${}^{203}$Tl(p,3n)${}^{201}$Pb$\longrightarrow$ ${}^{201}$Tl & Sakai (Sakai et al., 1965)\newline Lebowitz (Lebowitz et al., 1975)\newline Blue (Blue et al., 1978)\newline Lagunas-Solar (Lagunas-Solar et al., 1978)\newline Qaim (Qaim et al., 1979)\newline Bonardi (Birattari et al., 1982)\newline Hermanne (Hermanne et al., 1992)\newline Al-Saleh (Al-Saleh et al., 2007) & Only  presently  used in practice \\ \hline 
${}^{205}$Tl(p,xn)${}^{201}$Pb$\longrightarrow$${}^{201}$Tl & Lagunas-Solar (Lagunas-Solar et al., 1980) & Requires high energy beam \\ \hline 
${}^{203}$Tl(d,4n)${}^{201}$Pb$\longrightarrow$${}^{201}$Tl & Adam Rebles (Adam-Rebeles et al., 2012)\newline Blue (Blue et al., 1978) & Requires high energy beam \\ \hline 
${}^{201}$Hg(p,n)${}^{201}$Tl & Fernandes (Fernandes and da Silva, 1992)  \newline Comar (Comar and Crouzel, 1975) & Low cross-section \\ \hline 
${}^{202}$Hg(p,2n)${}^{ 201}$Tl & Birattari (Birattari et al., 1982) & ${}^{202}$Tl impurity level \\ \hline 
${}^{200}$Hg(d,n)${}^{ 201}$Tl\newline ${}^{201}$Hg(d,2n)${}^{ 201}$Tl & Dmitriev (Dmitriev et al., 1976)\newline Comar (Comar and Crouzel, 1975) & Low cross-section\newline ${}^{202}$Tl impurity level \\ \hline 
${}^{nat}$Pb(p,x)${}^{201}$Tl & Zaitseva (Zaitseva et al., 1987)\newline  Lagunas-Solar (Lagunas-Solar et al., 1981)\newline Gloris (Gloris et al., 2001; Kuhnhenn et al., 2001)\newline Bell (Bell and Skarsgard, 1956) & Requires high energy beam \\ \hline 
${}^{197}$Au(${}^{7}$Li,p2n)${}^{201}$Tl & Nayak (Nayak et al., 2002) & Easy targetry\newline Poor availability of ${}^{7}$Li beam \\ \hline 
${}^{197}$Au(${}^{6}$Li,pn)${}^{201}$Tl &  & Easy targetry\newline Poor availability of ${}^{6}$Li beam \\

\hline
\end{tabular}
\end{center}

\end{table*}

The final result of these studies and of the everyday practice is that the $^{203}Tl(p,3n)^{201}Pb\longrightarrow ^{201}Tl$ production route  appears to be the most practical one taking into account the targetry, the production yields, the available accelerators, the radionuclide purity, etc.  Presently the $^{203}Tl(p,3n)$ reaction is used exclusively for production of $^{201}Tl$. In the next chapters we are discussing only this production route together with side reactions during bombardment of thallium target with protons. 
In Table 2 we have collected the most important physical parameters of the experiments related to the measurement of activation cross-section and yield of proton induced nuclear reactions on thallium \citep{4,8,9,17,26,29,34,42,43,44,46,52,54,55}.
According to Table 2, a large number of papers was published on $Tl(p,x)^{201}Pb\longrightarrow ^{201}Tl$ production and chemistry, on compilation of the earlier experimental data and on the calculation of cross-section by using different theoretical model codes. In Table 3 we have summarized the most important compilations and theoretical calculations \citep{12,18,24,25,28,30,33,36,38,42,45,48,53,56,57,58,59,60,62}.
In spite of numerous earlier works we decided to make new measurements and new evaluations on the activation cross-sections of proton induced reactions on thallium.
The aim of the present study and the report was manifold:
\begin{enumerate}
\item 	Some part of the recently reported experimental data were measured long time ago \citep{29}. Only preliminary results were reported in equidistant energy scale obtained by fitting the experimental data, without any details on experiments and on the nuclear data used for the data evaluation. This fitted preliminary data were used for preparation of recommended data file of IAEA \citep{28} for production of $^{201}Tl$  and used in the upgraded data library \citep{60} without correction of  the decay data. In this work we decided to publish the original experimental data corrected with upgraded decay data for further use to prepare recommended data (see this work - ser. 1).
\item 	During the last decade new experiments were performed, which could be included into the new evaluation. It is necessary to check the earlier works in respect of new monitor and decay data.
\item 	The theoretical model codes were significantly improved in the last decade and new input parameter library was developed. It was a challenge to see the predictivity of these codes in comparison with the experimental data and with the earlier model results.
\item 	We have performed experiments on deuteron induced reactions on $^{nat}Tl$ up to 50 MeV \citep{1}. We thought it is worthwhile to repeat the experiments with proton beam using the same experimental technology (target preparation, spectra measurements, etc.).
\item 	Last but not least in connection with investigation of production possibility of $^{201}Tl$ in Hungary, the ATOMKI group had direct practical interest in the targetry and the production yields (similar to the case of other routinely produced isotopes ($^{11}C$, $^{18}F$, $^{67}Ga$, $^{111}In$, $^{123}I$).
In such a way, to complete our  systematical study of excitation functions for production of medical radioisotopes and to make new evaluation we decided to make new measurement relative to the well-established monitor reactions by using more modern experimental and evaluation techniques (see this work ser. 2).
\end{enumerate}

\begin{table*}[t]
\tiny
\caption{Summary of earlier experimental investigations on activation cross-sections and yields of the proton induced nuclear reaction on thallium (abbreviations for the measured quantities are taken from the EXFOR, see in footnote) }
\centering
\begin{center}
\begin{tabular}{|p{0.8in}|p{0.7in}|p{0.6in}|p{0.7in}|p{1.0in}|p{1.5in}|p{0.7in}|} 
\hline 
\textbf{Author} & \textbf{Target} & \textbf{Irradiation} & \textbf{Beam current\newline measurement\newline and \newline monitor reaction} & \textbf{Measurement \newline of activity\newline and\newline separation method} & \textbf{Nuclear reaction,\newline measured quantity,\newline number of measured data points } & \textbf{Covered energy\newline range\newline (MeV)} \\ \hline 
Sakai, & ${}^{nat}$Tl${}^{ }$ & cyclotron & Faraday cup & chemical separation & ${}^{nat}$Tl(p,x)${}^{200}$Pb, s, 6  & 10.5-51.6 \\  
(Sakai et al., 1965) &  & stacked foil &  & $\gamma$-Na(I) & ${}^{nat}$Tl(p,x)${}^{201}$Pb, s, 6 & 10.4-51.7 \\ 
 &  &  &  &  & ${}^{nat}$Tl(p,x)${}^{203}$Pb, s, 6 & 10.0-51.5 \\ \hline
Lebowitz, & ${}^{nat}$Tl & cyclotron & Faraday cup & chemical separation & ${}^{203}$Tl(p,3n)${}^{201}$Pb, s, 6  & 19.28-32.15 \\ 
(Lebowitz et al., 1975) & \~{}0.2 mg/cm${}^{2}$ & stacked foil &  & $\gamma$-Ge(Li) &  &  \\ \hline 
Milazzo-Colli & ${}^{203}$Tl, ${}^{205}$Tl & cyclotron &  & $\alpha$-Si surface barrier & ${}^{203}$Tl(p,a)${}^{200}$Hg, s, 1  & 20 \\ 
(Milazzo-Colli et al., 1975) & 300-400 $\mu $g/cm${}^{2}$ &  &  &  & ${}^{205}$Tl(p,a)${}^{202}$Hg, s, 1 & 20 \\ \hline 
Dmitriev, & ${}^{nat}$Tl & cyclotron & Faraday cup & no chemical separation & ${}^{203}$Tl(p,np)${}^{202}$Tl, TTY, 1 & 22 \\ 
(Dmitriev et al., 1976) & thick target & single foil &  & $\gamma$-Ge (Li) & ${}^{203}$Tl(p,3n)${}^{201g}$Pb,TTY, 1  & 22 \\  
 &  &  &  &  & ${}^{203}$Tl(p,2n)${}^{202m}$Pb, TTY, 1  & 22 \\  
 &  &  &  &  & ${}^{nat}$Tl(p,x)${}^{203g}$Pb, TTY, 1   & 22 \\ \hline
Hanna, & ${}^{nat}$Tl & cyclotron & Faraday cup & chemical separation & ${}^{nat}$Tl(p,x)${}^{200}$Pb,TTD, REL, 4 & 26.89-34.84 \\ 
(Hanna et al., 1977) & 275 mg/cm${}^{2}$ & EN tandem VdG &  & $\gamma$-Ge (Li) & ${}^{nat}$Tl(p,x)${}^{201}$Pb,TTD, REL, 6  & 20.4-34.87  \\  
 &  & stacked foil &  &  & ${}^{nat}$Tl(p,x)${}^{202}$Tl, TTD, REL, 7  & 16.27-34.84 \\ \hline
J. W. Blue, & ${}^{nat}$Tl foil & cyclotron & Faraday cup & no chemical separation & ${}^{203}$Tl(p,3n)${}^{201}$Pb, s, 12 & 18.47-35.19 \\
 (Blue et al., 1978) & 110$\mu m$ & stacked foil &  & $\gamma$-Ge(Li) & ${}^{203}$Tl(P,4N)${}^{200}$Pb, s,  6  & 27.91-35.1 \\ \hline 
Lagunas-Solar  & ${}^{nat}$Tl & cyclotron & Faraday cup & no chemical separation & ${}^{nat}$Tl(p,x)${}^{200}$Pb, PY, FCT, 31 & 28.4-59.3 \\ 
(Lagunas-Solar et al., 1978) & 0.16-0.20 g/cm${}^{2}$ & stacked foil &  & $\gamma$-Ge(Li) & ${}^{203}$Tl(p,x)${}^{200}$Pb, s, DERIV, 14  & 30.0-40.9 \\ 
  &  &  &  &  & ${}^{nat}$Tl(p,x)${}^{201}$Pb, PY, FCT, 68  & 17.2-59.0 \\  
 &  &  &  &  & ${}^{203}$Tl(p,x)${}^{201g}$Pb, M+, s, DERIV, 39  & 18.3-33.6 \\ 
 &  &  &  &  & ${}^{nat}$Tl(p,x)${}^{201}$Tl, TTY, DT, 41  & 11.8-35.9 \\  
 &  &  &  &  & ${}^{203}$Tl(p,x)${}^{202m}$Pb, s, DERIV, 41   & 17.0-27.4 \\ 
 &  &  &  &  & ${}^{nat}$Tl(p,x)${}^{203}$Pb, PY, FCT, 45  & 17.0-59.1 \\ 
 &  &  &  &  & ${}^{nat}$Tl(p,x)${}^{203}$Pb, s, DERIV, 45 & 17.0-59.1 \\ 
 &  &  &  &  & ${}^{nat}$Tl(p,x)${}^{204m}$Pb, PY, FCT, 30  & 13.6-35.8 \\ 
 &  &  &  &  & ${}^{nat}$Tl(p,x)${}^{204m}$Pb, s, DERIV, 30  & 15.3-35. \\ 
 &  &  &  &  & ${}^{205}$Tl(p,x)${}^{200}$Pb, s, DERIV, 11  & 46.8-59.3 \\ 
 &  &  &  &  & ${}^{205}$Tl(p,x)${}^{201g}$Pb, M+, s, DERIV, 23  & 38.2-59.0 \\ 
 &  &  &  &  & ${}^{205}$Tl(p,x)${}^{202m}$Pb, s, DERIV, 23  & 27.4-35.9 \\ \hline 
Qaim et al. & ${}^{nat}$Tl & cyclotron & ${}^{63}$Cu(p,2n)${}^{62}$Zn & no chemical separation & ${}^{205}$Tl(p,2n)${}^{204m}$Pb, s, 27 & 12.1-44.2 \\ 
(Qaim et al., 1979) & 50-125$\mu m$  & stacked foil & ${}^{63}$Cu(p,p2n)${}^{61}$Cu & $\gamma$-Ge(Li) & ${}^{nat}$Tl(p,x)${}^{203}$Pb, IND, s, 34  & 7.49-44.4 \\ 
 &  &  &  &  & ${}^{nat}$Tl(p,x)${}^{202m}$Pb, IND, s, 31   & 9.59-44.2 \\ 
 &  &  &  &  & ${}^{nat}$Tl(p,x)${}^{201}$Pb, IND, s, 25   & 17.2-44.1 \\ 
 &  &  &  &  & ${}^{nat}$Tl(p,x)${}^{200}$Pb, IND, s, 14   & 28.2-44.4 \\ 
 &  &  &  &  & ${}^{nat}$Tl(p,x)${}^{199}$Pb, IND, s, 5  & 38.6-44.3 \\ 
 &  &  &  &  & ${}^{nat}$Tl(p,x)${}^{202}$Tl, IND,  18 & 16.2-44.2 \\ 
 &  &  &  &  & ${}^{nat}$Tl(p,x)${}^{201}$Tl, IND, s, 11  & 19.8-42.8 \\ \hline 
Bonardi  & ${}^{nat}$Tl foil & cyclotron & Faraday cup & $\gamma$-Ge(Li) & ${}^{nat}$Tl(p,x)${}^{200}$Pb, TTD, 7  & 27.696-42.24 \\ 
 (Bonardi et al., 1982) &  & stacked foil &  &  & ${}^{nat}$Tl(p,x)${}^{201}$Pb, TTD, TM, 12  & 18.823-42.067 \\ 
 &  &  &  &  & ${}^{nat}$Tl(p,x)${}^{202}$Tl, TTD, TTY, TM, 13   & 16.677-42.143 \\  
 &  &  &  &  & ${}^{nat}$Tl(p,x)${}^{203}$Pb, TTD, TTY, TM, 17  & 7.3794-42.177 \\ 
 &  &  &  &  & ${}^{203}$Tl(p,x)${}^{200}$Pb, TTD, 6   & 28.234-36.021 \\ 
 &  &  &  &  & ${}^{203}$Tl(p,x)${}^{201}$Pb, TTD, 11   & 20.755-35.984 \\ 
 &  &  &  &  & ${}^{203}$Tl(p,x)${}^{202m}$Pb, TTD, 14   & 14.015-36.023 \\  
 &  &  &  &  & ${}^{203}$Tl(p,x)${}^{203}$Pb, TTD, 14 & 14.001-35.983 \\ 
 &  &  &  &  &  &  \\ \hline 
Kernert (Kernert et al., 1983) & ${}^{203}$Tl (99\%) & cyclotron & Faraday cup &  & ${}^{203}$Tl(p,x)${}^{201}$Pb, TTY,  1  & 29 \\  
 & ${}^{nat}$Tl & single target &  &  & ${}^{203}$Tl(p,x)${}^{202m}$Pb, TTY, 1   & 29 \\  
 & thick target &  &  &  & ${}^{203}$Tl(p,x)${}^{203}$Pb, TTY, 1   & 29 \\  
 &  &  &  &  & ${}^{203}$Tl(p,x)${}^{204m}$Pb, TTY, 1   & 29 \\  
 &  &  &  &  & ${}^{nat}$Tl(p,x)${}^{201}$Pb, TTY, 1   & 29 \\ 
 &  &  &  &  & ${}^{nat}$Tl(p,x)${}^{202m}$Pb, TTY, 1   & 29 \\  
 &  &  &  &  & ${}^{nat}$Tl(p,x)${}^{203}$Pb, TTY, 1   & 29 \\ 
 &  &  &  &  & ${}^{nat}$Tl(p,x)${}^{204m}$Pb, TTY, 1   & 29 \\ \hline 
Malinin & ${}^{203}$Tl (96.9\%) & cyclotron & Faraday cup & chemical separation & ${}^{203}$Tl(p,x)${}^{201}$Pb, TTY, 1  & 30.5 \\  
(Malinin et al., 1984) & 0.35-0.4 g/cm${}^{2}$ & single target &  & $\gamma$-Ge(Li) &  &  \\ \hline 
 
Hermanne  & ${}^{nat}$Tl (50$\mu$ m)  & cyclotron & Faraday cup & no chemical separation & ${}^{203}$Tl(p,2n)${}^{202m}$Pb, s,  DERIV, 13 & 15-27 \\ 
(Hermanne et al., 1992) &  & stacked foil &  & $\gamma$-Ge(Li) & ${}^{203}$Tl(p,3n)${}^{201}$Pb, s, DERIV, 22  & 19-42 \\  
 &  &  &  &  & ${}^{203}$Tl(P,4N)${}^{200}$Pb, s,  DERIV, 12   & 29-42 \\  
 &  &  &  &  & ${}^{205}$Tl(p,2n)${}^{204m}$Pb, s, DERIV, 29  & 12-42 \\ 
 &  &  &  &  & ${}^{203}$Tl(p,3n)${}^{203}$Pb, s, DERIV, 24  & 17-40 \\ 
 &  &  &  &  & ${}^{203}$Tl(p,4n)${}^{202m}$Pb, s, DERIV, 12  & 29-40 \\ 
 &  &  &  &  & ${}^{203}$Tl(p,2n)${}^{201}$Pb, s, DERIV, 4 & 38-42 \\ \hline 
Sattari & ${}^{203}$Tl(enr.) & cyclotron & Faraday cup & $\gamma$-HpGe(Li) & ${}^{203}$Tl(p,x)${}^{201}$Pb, TTY,  REL, 4 & 28-30 \\ 
(Sattari et al., 2003) & 18.3 $\mu$ m & single target &  &  &  &  \\ \hline 
Al-Saleh & ${}^{nat}$Tl & cyclotron & Faraday cup & no chemical separation & ${}^{nat}$Tl(p,x)${}^{201}$Pb, s = ${}^{203}$Tl(p,3n)${}^{201}$Pb, s, 17  & 19.3-27.2 \\ 
(Al-Saleh et al., 2007) & 10-15$\mu$ m thick on 10$\mu$ m Cu & stacked foil & ${}^{29}$Cu(p,x)${}^{63}$Zn monitor\newline  & g-HpGe & ${}^{nat}$Tl(p,x)${}^{202m}$Pb, s = ${}^{203}$Tl(p,2n)${}^{202m}$Pb, s + ${}^{205}$Tl(p,4n)${}^{202m}$Pb, s, 23  & 15.7-27.2 \\ 
 &  &  &  &  & ${}^{nat}$Tl(p,x)${}^{203}$Pb, s = ${}^{203}$Tl(p,n)${}^{203}$Pb, s + ${}^{205}$Tl(p,3n)${}^{203}$Pb, s, 34 & 6.2-27.2 \\ \hline 
 &  &  &  &  & ${}^{nat}$Tl(p,x)${}^{204m}$Pb, s = ${}^{205}$Tl(p,2n)${}^{204m}$Pb, s, 28 & 11.3-27.2 \\  
 &  &  &  &  &  &  \\ \hline 
this work Ser. 1 & ${}^{nat}$Tl (50$\mu$ m) & cyclotron & Faraday cup & no chemical separation & ${}^{nat}$Tl(p,x)${}^{204m}$Pb, 105 & 12.2-41.7 \\  
 & on & stacked foil &  & $\gamma$-Ge(Li) & ${}^{nat}$Tl(p,x)${}^{203}$Pb, 73 & 10.8-41.7 \\ 
 & Cu (50$\mu$ m)  &  &  &  & ${}^{nat}$Tl(p,x)${}^{202m}$Pb, N110 &  \\ 
 &  &  &  &  & ${}^{nat}$Tl(p,x)${}^{201}$Pb, 81 &  \\ 
 &  &  &  &  & ${}^{nat}$Tl(p,x)${}^{200}$Pb, 67 & 10.8-41.7 \\ 
 &  &  &  &  & ${}^{nat}$Tl(p,x)${}^{199}$Pb, 17 & 17.1-41.7 \\ \hline 
this work Ser. 2 & ${}^{nat}$Tl (20.8 $\mu$ m) & cyclotron & Faraday cup & no chemical separation & ${}^{nat}$Tl(p,x)${}^{204m}$Pb, 20 & 27.1-41.5 \\ 
 & on  &  & ${}^{nat}$Cu(p,x)${}^{62,65}$Zn & $\gamma$-HpGe & ${}^{nat}$Tl(p,x)${}^{203}$Pb, 20 &  \\ 
 & Cu (12.5 $\mu$ m) &  &  &  & ${}^{nat}$Tl(p,x)${}^{202m}$Pb, 20 &  \\  
 &  &  &  &  & ${}^{nat}$Tl(p,x)${}^{201}$Pb, 20 &  \\ 
 &  &  &  &  & ${}^{nat}$Tl(p,x)${}^{200}$Pb, 20 &  \\ 
 &  &  &  &  & ${}^{nat}$Tl(p,x)${}^{202}$Tl, 20 &  \\ 
 &  &  &  &  & ${}^{nat}$Tl(p,x)${}^{201}$Tl, 20 &  \\ 
 &  &  &  &  & ${}^{nat}$Tl(p,x)${}^{200}$Tl, 20 &  \\ 
 &  &  &  &  & ${}^{nat}$Tl(p,x)${}^{203}$Hg, 20 &  \\ 

\hline
\end{tabular}
\end{center}
\begin{flushleft}
\footnotesize{SIG-Cross section, TTY-thick target yield, TTD-differential thick target yield, DERIV-derived data, IND-independent formation, CUM-cumulative formation, REL-relative
}
\end{flushleft}
\end{table*}

\begin{table*}[t]
\tiny
\caption{The most important compilations and theoretical calculations on cross-sections and yields of ${}^{nat}$Tl(p,x)${}^{201}$Pb?${}^{201}$Tl and the side reactions}
\centering
\begin{center}
\begin{tabular}{|p{1.2in}|p{1.2in}|p{1.2in}|p{1.2in}|p{1.2in}|} 
\hline 

\textbf{Author} & \textbf{Model code} & \textbf{Reactions} & \textbf{Compilation} & \textbf{Recommended} \\ \hline 
Canedrias Ruz  (Canderias-Cruz and Okamoto, 1987) &  & ${}^{nat}$Tl(p,x)${}^{201}$Pb & X &  \\ \hline 
Rurarz (Rurarz, 1994) &  & ${}^{nat}$Tl(p,x)${}^{201}$Pb & X &  \\ \hline 
Kurenkov (Kurenkov et al., 1995) & ALICE-87, STAPRE & ${}^{203,205}$Tl(p,xn) &  &  \\ \hline 
IAEA TECDOC (Hermanne et al., 2001) & ALICE-IPPE, ALICE-HMS & ${}^{nat}$Tl(p,x)${}^{201}$Pb\newline ${}^{nat}$Tl(p,x)${}^{202}$Pb\newline ${}^{nat}$Tl(p,x)${}^{200}$Pb & X & X \\ \hline 
Takács (Takács et al., 2005) &  & ${}^{nat}$Tl(p,x)${}^{201}$Pb\newline ${}^{nat}$Tl(p,x)${}^{202}$Pb\newline ${}^{nat}$Tl(p,x)${}^{200}$Pb & X & X \\ \hline 
Tel (Tel et al., 2011) & ALICE/ASH & ${}^{nat}$Tl(p,x)${}^{201}$Pb &  &  \\ \hline 
Haji Said (Haji-Saeid et al., 2009) &  & ${}^{nat}$Tl(p,x)${}^{201}$Pb & X &  \\ \hline 
Sheu  (Sheu et al., 2003) & ALICE-01, Fluka & ${}^{nat}$Tl(p,x)${}^{201}$Pb & X &  \\ \hline 
Landolt Bornstein (Iljinov et al., 1993) &  & ${}^{nat}$Tl(p,x)${}^{201}$Pb\newline ${}^{nat}$Tl(p,x)${}^{202}$Pb\newline ${}^{nat}$Tl(p,x)${}^{200}$Pb & X &  \\ \hline 
Mihãilescu (Mihãilescu et al., 2007) &  &  &  &  \\ \hline 
Lagunas-Solar (Lagunas-Solar et al., 1978) & STAPRE &  &  &  \\ \hline 
Groppi (Groppi et al., 2001) &  & ${}^{nat}$Tl(p,x)${}^{201}$Pb & X &  \\ \hline 
Shubin (Shubin, 2001) &  &  &  &  \\ \hline 
Nowotny (Nowotny, 1981) &  &  &  &  \\ \hline 
MENDL -2P (Shubin et al., 1998) & ALICE-IPPE & ${}^{nat}$Tl(p,x)${}^{201}$Pb\newline ${}^{nat}$Tl(p,x)${}^{202}$Pb\newline ${}^{nat}$Tl(p,x)${}^{200}$Pb &  &  \\ \hline 
Koning (Koning and Rochman, 2011) & TALYS & ${}^{nat}$Tl(p,x)${}^{201}$Pb\newline ${}^{nat}$Tl(p,x)${}^{202}$Pb\newline ${}^{nat}$Tl(p,x)${}^{200}$Pb &  &  \\ \hline 
Szelecs\'enyi (Szelecs\'enyi et al., 1995) &  &  &  &  \\ \hline 
Dmitriev (Dmitriev and Zaitseva, 1996) &  &  & X &  \\ \hline 
Kaplan \newline (Kaplan et al., 2009) & CEM cascade-exciton model\newline GDH model\newline Hybrid model \newline Equilibrium model & ${}^{203}$Tl(p,n)${}^{203}$Pb\newline ${}^{203}$Tl(p,2n)${}^{ 202}$Pb\newline ${}^{203}$Tl(p,3n)${}^{ 201}$Pb\newline ${}^{203}$Tl(p,4n)${}^{ 200}$Pb\newline ${}^{205}$Tl(p,3n)${}^{ 203}$Pb\newline ${}^{205}$Tl(p,4n)${}^{ 202}$Pb\newline ${}^{205}$Tl(p,5n)${}^{ 201}$Pb\newline ${}^{205}$Tl(p,6n)${}^{ 200}$Pb &  &  \\ \hline 

\hline
\end{tabular}
\end{center}
\end{table*}

\begin{table*}[t]
\tiny
\caption{Comparison of the $\gamma$-ray intensities used by different authors}
\centering
\begin{center}
\begin{tabular}{|p{0.5in}|p{0.6in}|p{0.6in}|p{0.5in}|p{0.4in}|p{0.4in}|p{0.6in}|p{0.6in}|p{0.6in}|p{0.6in}|} 
\hline 

\textbf{Nuclide}\textbf{J${}^{p}$}\newline \textbf{level energy (MeV)} & \textbf{Half-life} & \textbf{E$_\gamma$(keV)}\newline \textbf{NUDAT} & \textbf{I$_\gamma$ (\%)}\newline \textbf{NUDAT} & \textbf{I$_\gamma$(\%)}\newline \textbf{Lag-Sol} & \textbf{I$_\gamma$ (\%)}  \textbf{Qaim} & \textbf{I$_\gamma$ (\%)} \textbf{Al Saleh} & \textbf{I$_\gamma$ (\%)}  \textbf{Bonardi} & \textbf{I$_\gamma$ (\%)} \textbf{Dmitriev} & \textbf{I$_\gamma$ (\%)}\newline \textbf{Hermanne 1991} \\ \hline 
${}^{204m}$Pb\newline 9-\newline 2185.88 & 66.93 m & 374.76\newline 899.15\newline 911.74 & 94.20\newline 99.174\newline 91.5 &  99.2 & 95\newline 100\newline 100\newline  & 89\newline 99 &  &  & 94.2\newline 99.3 \\ \hline 
${}^{203}$Pb\newline 5/2- & 51.92 h & 279.1952\newline 401.320 & 80.9\newline 3.35 & 81.0 & 80.7 & 99\newline 90.7 & 80.8\newline 3.35 & 81\newline 3.8 & 80\newline 3.8 \\ \hline 
${}^{202m}$Pb\newline 9-\newline 2169.83 & 3.54 h & 422.12\newline 657.49\newline 786.99\newline 960.70\newline 389.94\newline 459.72\newline 490.47 & 84\newline 31.7\newline 49\newline 89.9\newline 6.6\newline 9.2\newline ~9.8 & 85.3 & 90\newline \newline 56\newline 89 & 96\newline \newline 50 & 85.66\newline 32.4\newline 49.8\newline 91.3 &  & 86\newline 35\newline 50\newline 89 \\ \hline 
${}^{201}$Pb\newline 5/2-${}^{+}$ & 9.33 h & 331.15\newline 361.25\newline 405.96\newline 584.60\newline ~692.41\newline ~767.26\newline 826.26\newline 907.67\newline ~945.96 & 77\newline 9.5\newline 2.03\newline 3.6\newline 4.3\newline 3.28\newline 2.38\newline 6.1\newline 7.2 & 61.4 & 82\newline 10.7 & 79\newline 9.9 &  3.56\newline 4.27\newline \newline \newline 5.70\newline 7.36 &  & 82\newline 10.7 \\ \hline 
${}^{200}$Pb\newline 0+ & 21.5 h & 147.63\newline 235.62\newline 257.19\newline 268.36\newline 450.56 & 38.2\newline 4.35\newline 4.52\newline 4.01\newline 3.37 & 28.4 & 30.8 &  & 37.73\newline 4.3\newline 4.46\newline 3.96 &  & 29 \\ \hline 
${}^{199}$Pb\newline 3/2- & 90 m & 366.90\newline 720.24\newline 1135.04 & 44 \%\newline 6.5\newline 7.8 &  &  11.6\newline 14 &  &  &  & 79\newline 11.4\newline  \\ \hline 
${}^{202}$Tl\newline 2- & 12.31 d & 439.510 & 91.5 &  & 95 & 91.4 &  &  &  \\ \hline 
${}^{201}$Tl\newline 1/2+ & 3.0421 d & 135.34\newline 167.43 & ~2.565\newline 10.00 &  & 2.65 & 2.565 &  &  &  \\ \hline 
${}^{200}$Tl\newline 2- & 26.1 h & 367.942\newline 579.300\newline 828.27\newline 1205.75 & 87\newline 13.7\newline 10.8\newline 30 &  &  & 87.2\newline 13.78\newline 10.81\newline 29.91 &  &  &  \\ \hline 
${}^{203}$Hg\newline 5/2- & 46.594 d & 279.1952 & 81.56 &  &  &  &  &  &  \\

\hline
\end{tabular}
\end{center}
\end{table*}

\begin{table*}[t]
\tiny
\caption{Nuclear data of the investigated reactions for ${}^{nat}$Tl+ p}
\centering
\begin{center}
\begin{tabular}{|p{0.5in}|p{0.6in}|p{0.6in}|p{0.6in}|p{0.5in}|p{0.7in}|p{0.8in}|} 
\hline 

\textbf{Nuclide}\newline \textbf{J${}^{p}$}\newline \textbf{level energy (MeV)} & \textbf{Half-life} & \textbf{Decay mode (\%)} & \textbf{E$_\gamma$\newline (keV)} & \textbf{I$_\gamma$ (\%)} & \textbf{Contributing} \newline \textbf{reactions} & \textbf{Q-value\ (keV)} \\ \hline 
${}^{204m}$Pb\newline 9-\newline 2185.88 & 66.93 m & IT(100) & 374.76\newline 899.15\newline 911.74 & 94.20\newline 99.174\newline 91.5 & ${}^{203}$Tl(p,g)\newline ${}^{205}$Tl(p,2n) & 6637.51\newline -7564.517 \\ \hline 
${}^{203}$Pb\newline 5/2- & 51.92 h & EC(100) & 279.1952\newline 401.320 & 80.9\newline 3.35 & ${}^{203}$Tl(p,n)\newline ${}^{205}$Tl(p,3n) & -1756.97\newline -15959.0 \\ \hline 
${}^{202m}$Pb\newline 9-\newline 2169.83 & 3.54 h & IT(90.5)\newline EC(9.5) & 422.12\newline 657.49\newline 786.99\newline 960.70\newline 389.94\newline 459.72\newline 490.47 & 84\newline 31.7\newline 49\newline 89.9\newline 6.6\newline 9.2\newline ~9.8 & ${}^{203}$Tl(p,2n)\newline ${}^{205}$Tl(p,4n) & -8681.26\newline -22883.29 \\ \hline 
${}^{201}$Pb\newline 5/2-${}^{+}$ & 9.33 h & EC(100) & 331.15\newline 361.25\newline 405.96\newline 584.60\newline ~692.41\newline ~767.26\newline 826.26\newline 907.67\newline ~945.96 & 77\newline 9.5\newline 2.03\newline 3.6\newline 4.3\newline 3.28\newline 2.38\newline 6.1\newline 7.2 & ${}^{203}$Tl(p,3n)\newline ${}^{205}$Tl(p,5n) & -17428.3\newline -31630.3 \\ \hline 
${}^{200}$Pb\newline 0+ & 21.5 h & EC(100) & 147.63\newline 235.62\newline 257.19\newline 268.36\newline 450.56 & 38.2\newline 4.35\newline 4.52\newline 4.01\newline 3.37 & ${}^{203}$Tl(p,4n)\newline ${}^{205}$Tl(p,6n) & 24514.2\newline -38716.2 \\ \hline 
${}^{199}$Pb\newline 3/2- & 90 m & EC(100) & 366.90\newline 720.24\newline 1135.04 & 44 \%\newline 6.5\newline 7.8 & ${}^{203}$Tl(p,5n)\newline ${}^{205}$Tl(p,7n) & -33600.8\newline -47802.9 \\ \hline 
${}^{202}$Tl\newline 2- & 12.31 d & EC(100) & 439.510 & 91.5 & ${}^{203}$Tl(p,pn)\newline ${}^{205}$Tl(p,p3n)\newline ${}^{202m}$Pb decay & -7849.2\newline -22051.3 \\ \hline 
${}^{201}$Tl\newline 1/2+ & 3.0421 d & EC(100) & 135.34\newline 167.43 & ~2.565\newline 10.00 & ${}^{203}$Tl(p,p2n)\newline ${}^{205}$Tl(p,p4n)\newline ${}^{201}$Pb decay & -14721.8\newline -28923.8 \\ \hline 
${}^{200}$Tl\newline 2- & 26.1 h & EC(100) & 367.942\newline 579.300\newline 828.27\newline 1205.75 & 87\newline 13.7\newline 10.8\newline 30 & ${}^{203}$Tl(p,p3n)\newline ${}^{205}$Tl(p,p5n)\newline  & -22927.04\newline -37129.08 \\ \hline 
${}^{203}$Hg\newline 5/2- & 46.594 d & $\beta $${}^{-}$(100) & 279.1952 & 81.56 & ${}^{205}$Tl(p,2pn) & -13911.76 \\

\hline
\end{tabular}
\end{center}
\begin{flushleft}
\footnotesize{Abundances (\%): $^{203}Tl$ (29.524), $^{205}Tl$ (70.476) 
Increase Q-values if compound particles are emitted: np-d, +2.2 MeV; 2np-t, +8.48 MeV; n2p-$^3$He, +7.72 MeV; 2n2p-$\alpha$, +28.30 MeV
Decrease Q-values for isomeric states with level energy of the isomer
}
\end{flushleft}
\end{table*}

\section{Experimental}
\label{2}

We have performed two series of experiments. The first series was measured 20 years ago and the second was done in the last year.

\subsection{Experimental and data evaluation details of the first series of measurements:}
\label{2.1}

Samples of natural Tl (50 $\mu m$) were prepared by electroplating on brass-foils (50 $\mu m$). No additional monitoring foils were inserted into the stacks. The irradiation was done at the VUB CGE 560 cyclotron at beam current of 3-6  $\mu A$, for 3-6 minutes. 13 stacks of 7-12 foils were irradiated at incident energies of 43-22 MeV. The number of incident particles was determined from the integrated charge measured in Faraday cup. Tl produced in irradiated samples was quantitatively dissolved after irradiation and aliquots of 1 ml sample were measured using a Ge(Li) detector calibrated with reference sources of $^{22}Na$, $^{152}Eu$, $^{60}Co$ (2\% error on standards). 
The proton energy degradation along the stack was computed using the stopping formulae coefficients of Andersen and Ziegler for Tl \citep{5} and the tabulated stopping powers of \citep{32}  for the brass backings.
The production cross-section was calculated from activity. Quadratic summation of the uncertainties of the contributing parameters results an absolute error of 8.5\% (target thickness (2\%), detector efficiency (5\%), physical decay data (5\%), integrated beam current (2\%), counting statistics (2\%)). 
Nuclear decay data originally were taken from Erdtmann and Soyka \citep{19}.  For preparation of the IAEA recommended data base  \citep{60} and for compilation into the EXFOR  format the decay data were corrected according to the newest parameters \citep{11}. In the recent work the cross-section data were corrected according to the new decay data, and the uncertainty of the number of incident particle on the basis of comparisons of the monitor reactions and the Faraday cup data was enlarged from 2 \% to 7 \%, resulting in 11 \% total uncertainty.
The results of the earlier data evaluation were corrected for the recent recommended data of the  $\gamma$-ray intensities. No corrections were done for the half-lives (nonlinearly contributing factor), but there are only small changes in half-lives. In a few cases there are significant changes in  $\gamma$-ray intensities (see Table 4). There are significant changes in decay data for $^{201}Pb$, $^{200}Pb$ used by Lagunas Solar et al., for 401 keV  $\gamma$-line of $^{203}Pb$ used by Al Saleh et al. \citep{4}. In the case of the $^{199}Pb$ the recent decay data also differ significantly from data used by Hermanne et al. \citep{29} and by Qaim et al. \citep{52}.
The correction for the decay data is complicated when the authors used many  $\gamma$-lines, for which different corrections are required according to the recent data. In this case the correction was done for the strongest  $\gamma$-line. 
It should be mentioned that the automatic corrections of the of the cross-section and yield data for the new decay data is not a simple question in the case of many used  $\gamma$-lines, taking into account the lack of the information in the original publications on the averaging of cross-section data obtained from different  $\gamma$-lines, measured at different time, sometimes at different sample - detector distances.

\subsection{Experimental  and data evaluation details of the second series of measurement:}
\label{2.2}

Samples of natural Tl were prepared also by electroplating.  Simultaneous plating of four 15  $\mu m$ thallium target layers (surface area 11.69 $cm^2$) on 12.5 $\mu m$ Cu foil were used. The plating method is described in detail in \citep{3}. The electroplated foils for preparation of target stack were cut for 10x10 mm pieces.
No additional monitoring foils were inserted into the stacks, the nuclear reactions induced on the backing Cu foils were used as beam monitors.
 The irradiation was done at VUB CGE 560 cyclotron in a short Faraday cup at beam current 90 nA, for 75 minutes irradiation for time, at 37 MeV incident energy. The radioactivity of the irradiated sample and monitor foils was measured nondestructively using HPGe  $\gamma$-spectrometry. Counting was started about 3-4 hours after the end of the bombardment (EOB). 
The proton energy degradation along the stack was computed using the stopping formulae coefficients of Andersen and Ziegler for Tl \citep{5}, and corrected on the basis of a simultaneously measured $^{nat}Cu(p,x)^{62}Zn,^{65}Zn$ monitor reactions \citep{61}.The uncertainties in the energy scale were estimated taking into account the uncertainty of the energy of the primary beam, the calculated beam straggling and the measures necessary for the energy corrections on the basis of simultaneously measured excitation functions of monitor reactions.

\begin{figure}[h]
\includegraphics[scale=0.3]{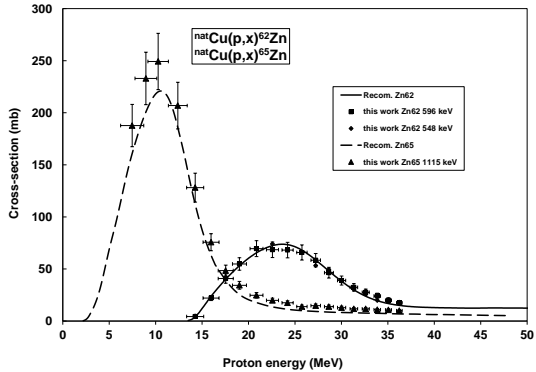}
\caption{Application of monitor reactions for determination of proton beam energy and intensity}
\end{figure}

The intensity of the incident beam was obtained from the charge collected in a Faraday cup and corrected using the monitor reactions. The best agreement was found after small corrections on the beam intensity (7\% with respect to the Faraday cup results) and on the primary beam energy (0.5 MeV). According to the Fig. 2 the agreement is good in the whole energy range.
The production cross-section was calculated from activity. So called "elemental" cross-sections were determined, considering that the Tl targets are monoisotopic. The decay data were taken from NuDat database \citep{49} the reaction Q-values from \citep{50} . The used decay data are summarized in Table 5. 
The uncertainties were estimated in the standard way \citep{31} by combining the individual errors in quadrature. The main sources of error were: beam current (8\%), counting statistics and peak area determination (1-7\%), detector efficiency (5\%), decay data (3\%) and effective target thickness (5\%)). A total uncertainty of 12-20\% is obtained.

\section{Theoretical calculations}
\label{3}

The measured cross-sections were compared to theoretical effective cross-sections calculated by using the ALICE-IPPE \citep{16} and EMPIRE-II  \citep{27}( codes as well as with the data based on the latest version of the TALYS code \citep{35}, retrieved from the online TENDL 2011 file \citep{36}.
An a priori calculation was done without any knowledge of the experimental results and thus without any parameter adjustment. For each activation product the reaction cross-section on the individual target isotopes was calculated and a weighted summation (according to the abundance of natural occurrence) was made to obtain the total production cross-section. 

\section{Results}
\label{4}

\subsection{Excitation functions}
\label{4.1}

The cross-sections for all the reactions studied are shown in Figures 3 - 13 and the numerical values are collected in Tables 6 - 8.  The results of the two series of measurements are shown separately. Our results of the two series of measurements are in acceptable good agreement. In most cases the radioisotopes are formed via nuclear processes occurring on $^{203}Tl$ and $^{205}Tl$.  The cross-sections presented here refer to so called effective "elemental" cross-sections, calculated on hypothetic monoisotopic thallium target having number of target nuclei equal to the number of the sum of the numbers of $^{205}Tl$ and $^{207}Tl$ isotopes. We discuss the different reaction products individually.

\subsubsection{Activation cross-sections of lead radioisotopes}
\label{4.1.1}

The radioisotopes of lead are produced via (p,xn) reactions on $^{203}Tl$ and $^{205}Tl$. The earlier experimental activation data on Tl are related mainly to the production of radioisotopes of lead, and was measured mostly up to 40 MeV.

\textbf{$^{nat}Tl(p,xn)^{204m}Pb$ reaction}

This radioisotope has two states: a shorter-lived ($T_{1/2}$ = 66.93 m, $I^{\pi} = 9^{-}$) decaying by IT to the ground state. The only reaction leading to the formation of $^{204m}Pb$ from natural thallium is the $^{205}Tl(p,2n)$ reaction (if we neglect the small contribution from $^{203}Tl(p,\gamma )$). The calculations with various codes are compared with the present- and the literature experimental data are shown in Fig. 3. The experimental data show acceptable good agreement with the theory (except Lagunas Solar \citep{42} data). Systematic energy shift to the higher energies can be observed for experimental data of Al-Saleh \citep{4}. No TENDL calculated data are presented for production of $^{204m}Tl$, only for total production cross-section of $^{204}Tl$.

\begin{figure}[h]
\includegraphics[scale=0.3]{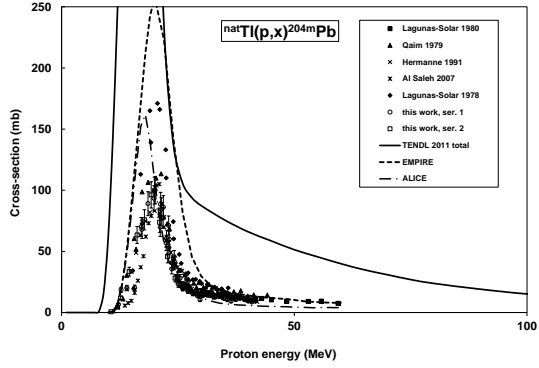}
\caption{Excitation function of $^{nat}Tl(p,x)^{204m}Pb$ reaction}
\end{figure}

\textbf{$^{nat}Tl(p,xn)^{203}Pb$ reaction}

The $^{203}Pb$ radioisotope is formed by $^{203}Tl(p,n)^{203}Pb$ and $^{205}Tl(p,3n)^{203}Pb$ reactions. The contributions of the reactions are well separated in Fig. 4. The available experimental data are in good agreement, except the data of Lagunas-Solar \citep{42} and Sakai \citep{54}   (see Fig. 4).  The theoretical data in TENDL 2011 describe well the experimental results.

\begin{figure}[h]
\includegraphics[scale=0.3]{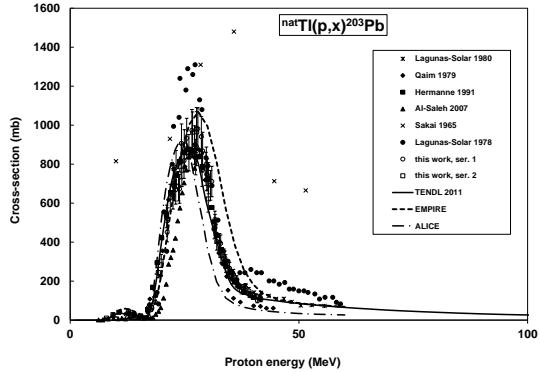}
\caption{Excitation function of $^{nat}Tl(p,x)^{203}Pb$ reaction}
\end{figure}

\textbf{$^{nat}Tl(p,xn)^{202m}Pb$ reaction}

The experimental and theoretical data for production of $^{202m}Pb$ are shown in Fig. 5. We were able to measure the production cross-section only of the shorter-lived metastable state (3.5 h), which decays mostly (90.5\%) by internal transition to the longer-lived (5250 a) ground state. In the investigated energy range both the $^{203}Tl(p,2n)$ and the $^{205}Tl(p,4n)$  contribute to the formation of $^{202m}Pb$ isomeric state. The comparison of the experimental data shows generally good agreement. At low energies the experimental data of Lagunas Solar \citep{42}  and \citep{4} shows energy shift towards high energies. There are no theoretical data in the TENDL 2011 for isomeric states, only for the total production of $^{202}Pb$.

\begin{figure}[h]
\includegraphics[scale=0.3]{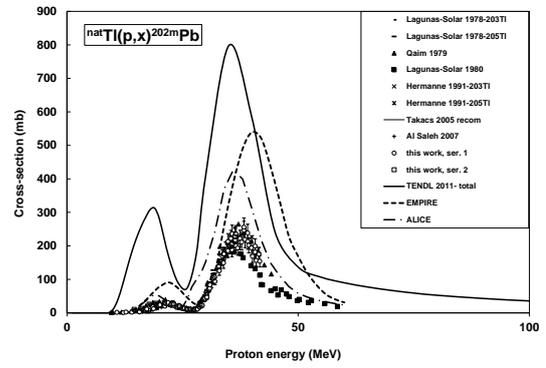}
\caption{Excitation function of $^{nat}Tl(p,x)^{202m}Pb$ reaction}
\end{figure}

\textbf{$^{nat}Tl(p,xn)^{201}Pb$ reaction}

The obtained new data are shown in Fig. 6 in comparison with the literature experimental data and with the theory. The structure of the excitation function (similar to previous processes) shows two maxima in the investigated energy range. The origin of the first peak is the $^{203}Tl(p,3n)^{201}Pb$ process. The second peak is due to the (p,5n) process on $^{205}Tl$. Our data in maximum is in good agreement with the results of Lagunas Solar \citep{42} , data  of Bonardi measured on enriched target \citep{10}, of Al-Saleh \citep{4} and Qaim \citep{52}. It should be mentioned that the data of Qaim \citep{52} are very scattered with approximately 30\% difference at the same energies (having 10-15\% uncertainties). Our data support the results of higher values (the low and high energy data came probably from beam current measurement of irradiations of different stacks). As it was mentioned at the previous reactions, the Al-Saleh data \citep{4} are shifted in energy near the threshold. The calculated cross-section curve follows the general trend of the experimental data.

\begin{figure}[h]
\includegraphics[scale=0.3]{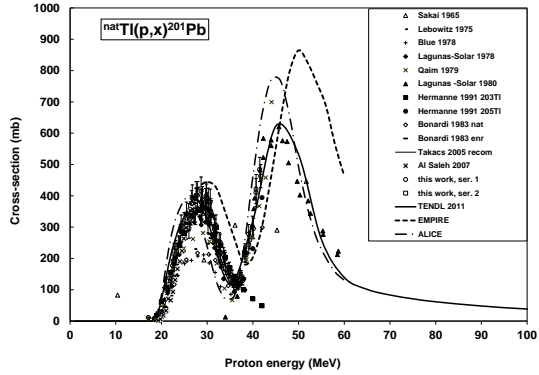}
\caption{Excitation function of $^{nat}Tl(p,x)^{201}Pb$ reaction}
\end{figure}

\textbf{$^{nat}Tl(p,xn)^{200}Pb$ reaction}

Irradiating thallium targets having natural isotopic composition the $^{203}Tl(p,4n)$ (Q = -24514.2 keV) and the $^{205}Tl(p,6n)$ (Q = -38716.2 keV)  reaction contribute to the production of $^{200}Pb$. According to Fig. 7 the agreement of experimental data is acceptable good, except the scattered data of Sakai et al \citep{54}, and the contradicting data of Lagunas-Solar at high energies reported in \citep{42} and his derived data \citep{41}. The theoretical data in TENDL 2011 describe well the low energy part where reliable experimental data exist.

\begin{figure}[h]
\includegraphics[scale=0.3]{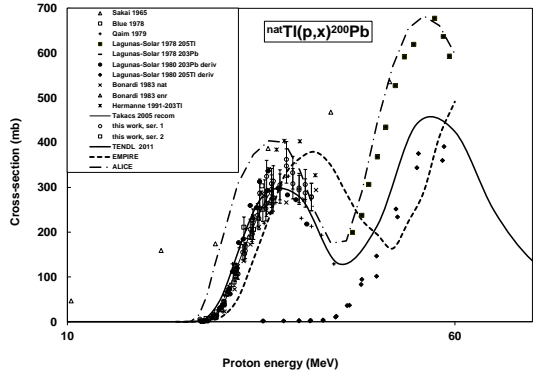}
\caption{Excitation function of $^{nat}Tl(p,x)^{200}Pb$ reaction}
\end{figure}

\textbf{$^{nat}Tl(p,xn)^{199}Pb$ reaction}

In the investigated energy range only the $^{203}Tl(p,5n)$ process contributes to the production of $^{199}Pb$. Only one earlier data set was found in the literature, reported by Qaim et al \citep{52}. Our new data show acceptable agreement with these experimental data and with the prediction of TALYS calculation in TENDL 2011 (see Fig. 8).

\begin{figure}[h]
\includegraphics[scale=0.3]{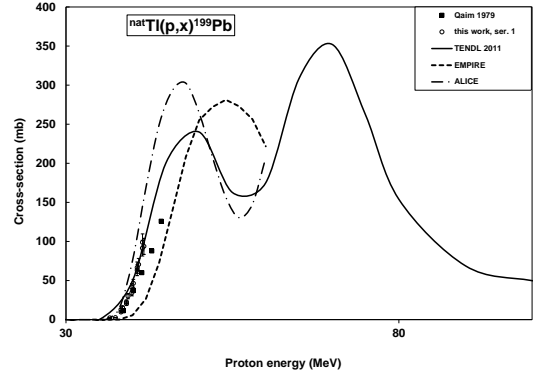}
\caption{Excitation function of $^{nat}Tl(p,x)^{199}Pb$ reaction}
\end{figure}

\subsubsection{Activation cross sections of thallium radioisotopes}
\label{4.1.2}

The radioisotopes of thallium are formed directly via (p,pxn) reaction  and through the decay of parent lead radioisotopes.

\textbf{$^{nat}Tl(p,xn)^{202}Tl$ reaction}

Parallel with the direct production the radioisotope is produced indirectly via the decay of the $^{202m}Pb$ ($T_{1/2}$ = 3.54 h, EC(9.5\%)) and of the decay of the long-lived $^{202g}Pb$ ($T_{1/2}$ = 52.5E+3 a, EC(100\%)). The cross-sections were calculated from spectra measured few days after EOB  i.e. after complete decay of $^{202m}Pb$, but insignificant decay of $^{202g}Pb$. In such a way our cumulative cross-sections contain contributions from the direct decay of $^{202m}Pb$. 
In Fig. 9 we show our cumulative data and corrected data for direct production of $^{202}Tl$  (when the contribution  from the $^{202m}Pb$ were subtracted). For comparison we also present the experimental data of Qaim \citep{52} corrected for the contribution of $^{202m}Pb$ decay (direct cross-sections), and  the theoretical data for the direct production.  The comparison shows good agreement with the earlier data and with theory.

\begin{figure}[h]
\includegraphics[scale=0.3]{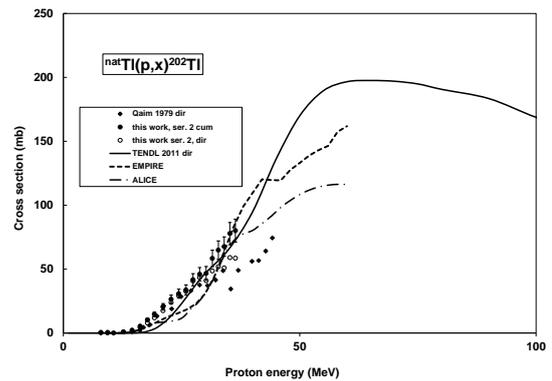}
\caption{Excitation function of $^{nat}Tl(p,x)^{202}Tl$ reaction}
\end{figure}

\textbf{$^{nat}Tl(p,xn)^{201}Tl$ reaction}

The $^{201}Tl$ ($T_{1/2}$ = 3.0421 d) radioisotope is formed directly through the $^{203}Tl(p,p2n)$, and $^{205}Tl(p,p4n)$ reactions and indirectly from the decay of $^{201}Pb$($T_{1/2}$ = 9.33 h). In Fig. 10 we present cumulative cross-section of $^{201}Tl$ after complete decay of $^{201}Pb$, and cross-section for direct production when the contribution from the decay of parent was subtracted. Our direct production cross-section data are systematically higher compared to the results of Qaim \citep{52}. The uncertainty of direct production is large, taking into account the large contribution of the parent. 

\begin{figure}[h]
\includegraphics[scale=0.3]{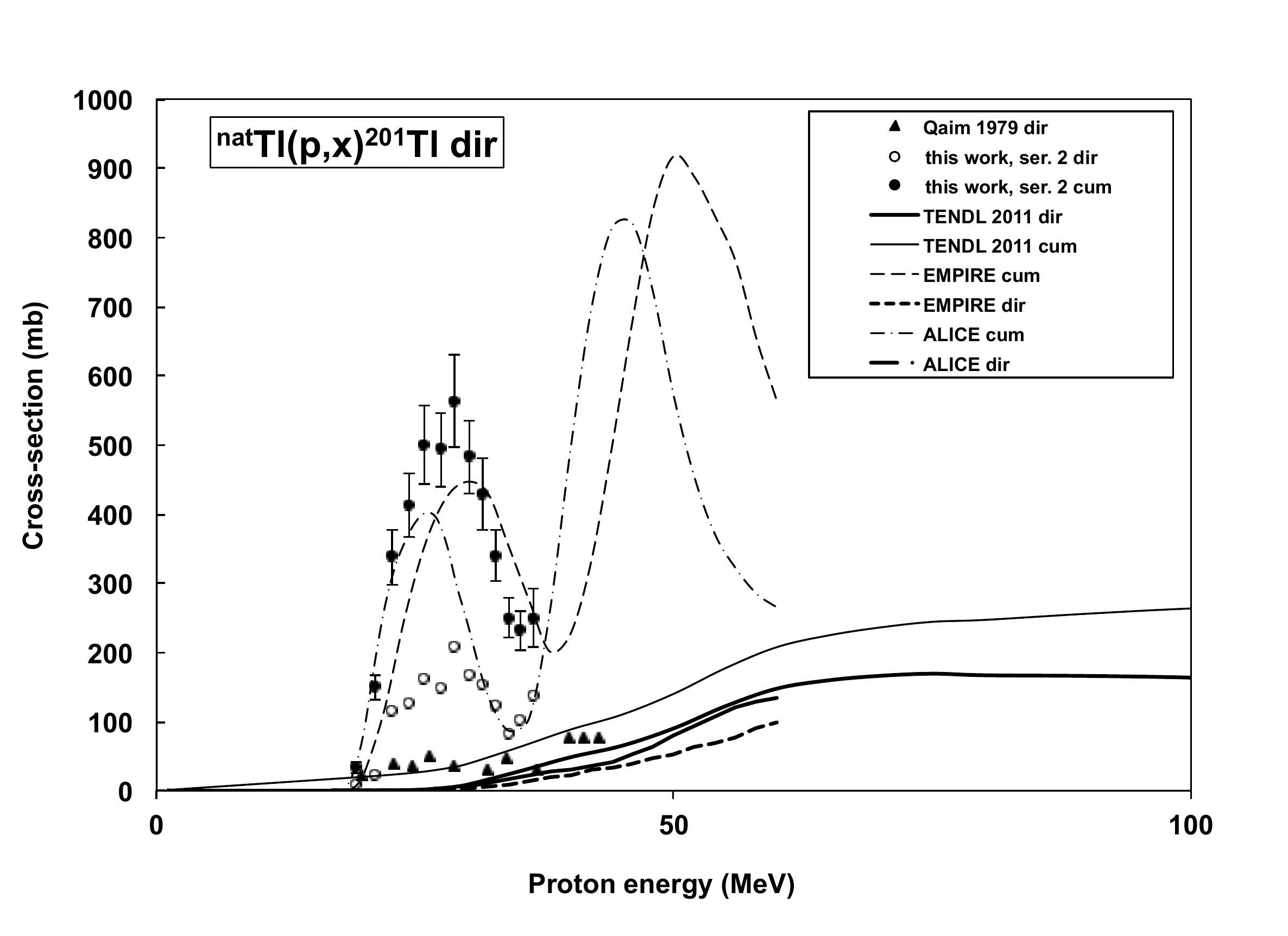}
\caption{Excitation function of $^{nat}Tl(p,x)^{201}Tl$ reaction}
\end{figure}

\textbf{$^{nat}Tl(p,xn)^{200}Tl$ reaction}

In Fig. 11 we present direct production cross-section for production of $^{200}Tl$ ($T_{1/2}$ = 26.1 h) after subtraction the contribution of the indirect production through the decay of $^{200}Pb$ ($T_{1/2}$ = 21.5 h). We have only a few points near the threshold and the uncertainties are high due to the high contribution of the parent. 

\begin{figure}[h]
\includegraphics[scale=0.3]{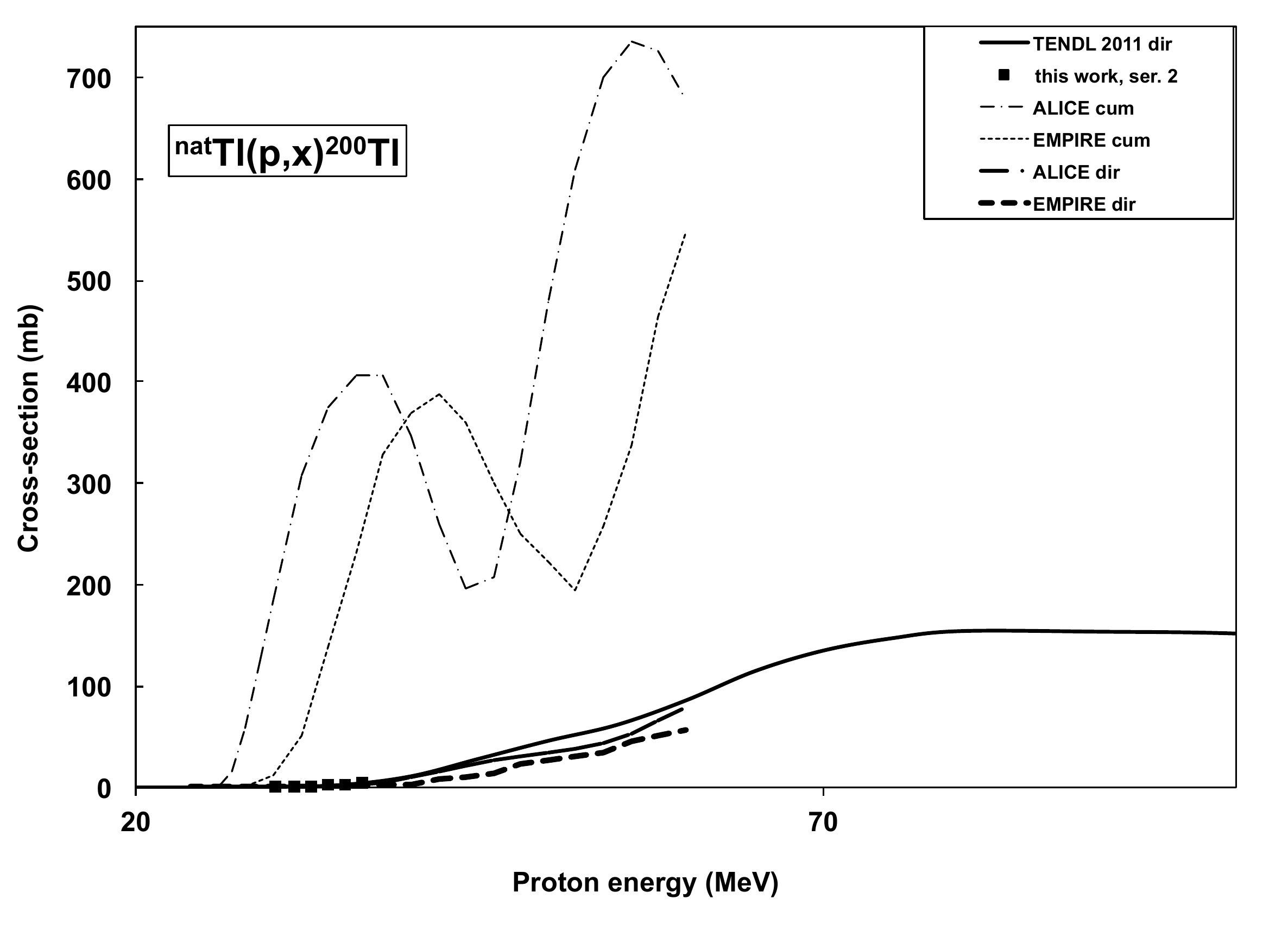}
\caption{Excitation function of $^{nat}Tl(p,x)^{201}Tl$ reaction}
\end{figure}

\subsubsection{Activation cross-sections of mercury radioisotopes}
\label{4.1.3}

The radioisotopes of mercury are formed directly via (p,2pxn) reaction  and through the decay of parent lead radioisotopes. We could measure activation cross-section only for production of $^{203}Pb$. In principle the thresholds for production of $^{199m}Hg$ (42.67 min, IT: 100\%), $^{197m}Hg$ (23.8 h, IT: 91.4\%) and $^{197g}Hg$ (64.14 h , EC 100\%) are lower than our maximum energy, but due to the low cross-sections (theoretical estimation in Fig. 12), to the weak  $\gamma$-lines  and  to the short half-life no reliable data were found.

\begin{figure}[h]
\includegraphics[scale=0.3]{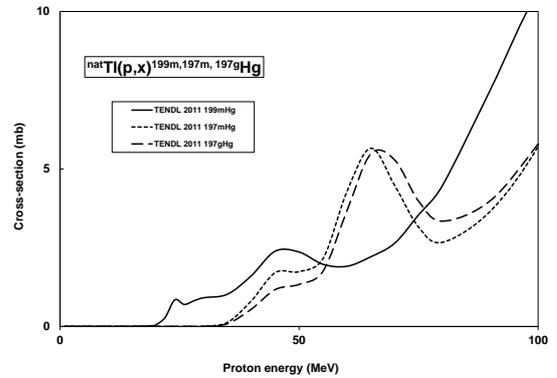}
\caption{Activation cross-sections for production of $^{199m}Hg$, $^{197m}Hg$ and $^{197g}Hg$ isomeric states predicted in TENDL 2011}
\end{figure}

\textbf{$^{nat}Tl(p,xn)^{203}Hg$ reaction}

The $^{203}Hg$ (46.594 d)  is produced through the $^{205}Tl(p,2pn)$ reaction. It has the same energy  $\gamma$-line (279 keV) as the $^{203}Pb$ (51.92 h), therefore it was measured after complete decay of $^{203}Pb$. Only a few experimental data points were measured above the effective threshold (Fig. 13). 

\begin{figure}[h]
\includegraphics[scale=0.3]{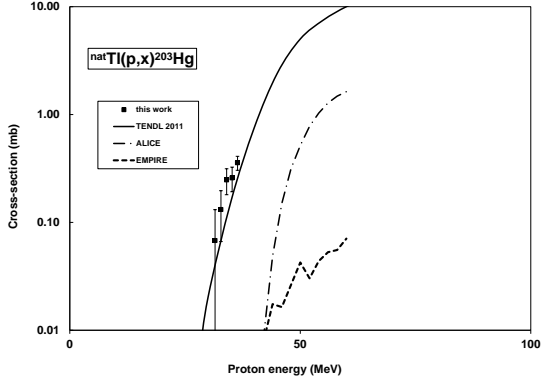}
\caption{Excitation function of $^{nat}Tl(p,x)^{203}Hg$ reaction}
\end{figure}

\begin{table*}[t]
\tiny
\caption{Measured cross-sections of the ${}^{204m}$Pb,${}^{ 203}$Pb,${}^{ 202m}$Pb, ${}^{201}$Pb and ${}^{ 199}$Pb reactions (ser. 1). Different cross-sections by the same energy come from different stack/irradiation}
\centering
\begin{center}
\begin{tabular}{|p{0.25in}|p{0.25in}|p{0.3in}|p{0.3in}|p{0.3in}|p{0.3in}|p{0.3in}|p{0.3in}|p{0.3in}|p{0.3in}|p{0.3in}|p{0.3in}|p{0.3in}|p{0.3in}|} 
\hline 
\multicolumn{2}{|p{0.5in}|}{\textbf{E $\pm$  $\Delta $E(MeV)}} & \multicolumn{12}{|p{3.6in}|}{\textbf{Cross-sections ($\sigma$) $\pm$ $\Delta\sigma$ (mb)}} \\ \hline 
\multicolumn{2}{|p{0.5in}|}{} & \multicolumn{2}{|p{0.6in}|}{${}^{204m}$Pb} & \multicolumn{2}{|p{0.6in}|}{ ${}^{203}$Pb} & \multicolumn{2}{|p{0.6in}|}{${}^{202m}$Pb } & \multicolumn{2}{|p{0.6in}|}{${}^{201}$Pb } & \multicolumn{2}{|p{0.6in}|}{${}^{200}$Pb} & \multicolumn{2}{|p{0.6in}|}{${}^{199}$Pb} \\ \hline 
10.8 & 2.5 &  &  & 41.5 & 4.6 & 1.3 & 0.15 &  &  &  &  &  &  \\ 
12.2 & 2.4 & 6.6 & 0.7 & 29.9 & 3.3 & 0.8 & 0.09 &  &  &  &  &  &  \\ 
12.8 & 2.3 & 18.5 & 2.0 & 47.3 & 5.2 & 2.5 & 0.27 &  &  &  &  &  &  \\ 
14.0 & 2.2 & 20.0 & 2.2 & 25.0 & 2.8 & 2.0 & 0.22 &  &  &  &  &  &  \\ 
15.7 & 2.1 & 20.2 & 2.2 & 9.8 & 1.1 & 2.9 & 0.32 &  &  &  &  &  &  \\ 
17.1 & 2.0 & 69.9 & 7.7 & 64.2 & 7.1 & 12.8 & 1.41 &  &  &  &  &  &  \\ 
17.3 & 2.0 & 67.9 & 7.5 &  &  & 11.8 & 1.30 & 12.2 & 1.34 &  &  &  &  \\  
18.6 & 1.9 & 88.9 & 9.8 & 92.1 & 10.1 & 18.1 & 1.99 &  &  &  &  &  &  \\  
18.7 & 1.9 &  &  &  &  & 12.6 & 1.39 &  &  &  &  &  &  \\  
19.9 & 1.8 & 93.3 & 10.3 & 300.6 & 33.1 & 25.1 & 2.76 &  &  &  &  &  &  \\ 
20.0 & 1.8 & 99.5 & 10.9 &  &  & 24.9 & 2.74 &  &  &  &  &  &  \\ 
20.1 & 1.8 & 99.9 & 11.0 &  &  & 24.3 & 2.67 &  &  &  &  &  &  \\ 
21.2 & 1.7 & 69.7 & 7.7 & 446.7 & 49.1 & 20.7 & 2.27 & 32.1 & 3.53 &  &  &  &  \\ 
21.3 & 1.7 & 88.4 & 9.7 & 514.1 & 56.6 & 29.6 & 3.26 & 101.7 & 11.18 &  &  &  &  \\ 
21.4 & 1.7 & 84.4 & 9.3 &  &  & 27.9 & 3.06 & 109.1 & 12.00 &  &  &  &  \\  
21.8 & 1.7 & 77.3 & 8.5 &  &  & 27.3 & 3.00 & 104.6 & 11.51 &  &  &  &  \\ 
21.9 & 1.7 & 85.9 & 9.5 & 621.6 & 68.4 & 31.5 & 3.46 & 146.6 & 16.12 &  &  &  &  \\ 
22.5 & 1.6 & 56.7 & 6.2 & 633.7 & 69.7 & 23.6 & 2.59 &  &  &  &  &  &  \\  
22.6 & 1.6 & 60.6 & 6.7 & 685.3 & 75.4 & 26.0 & 2.86 & 189.3 & 20.83 &  &  &  &  \\ 
23.0 & 1.6 & 71.0 & 7.8 &  &  & 30.7 & 3.38 & 202.7 & 22.29 &  &  &  &  \\ 
23.1 & 1.6 & 52.9 & 5.8 & 662.0 & 72.8 & 22.6 & 2.49 &  &  &  &  &  &  \\ 
23.1 & 1.6 & 59.8 & 6.6 & 776.6 & 85.4 & 27.0 & 2.97 & 247.2 & 27.19 &  &  &  &  \\  
23.7 & 1.6 & 35.5 & 3.9 & 680.1 & 74.8 & 16.6 & 1.83 &  &  &  &  &  &  \\  
23.8 & 1.6 & 40.7 & 4.5 &  &  & 20.4 & 2.24 &  &  &  &  &  &  \\ 
23.9 & 1.5 & 40.1 & 4.4 & 667.1 & 73.4 & 18.9 & 2.08 & 261.3 & 28.74 &  &  &  &  \\ 
24.2 & 1.5 & 45.6 & 5.0 &  &  & 22.4 & 2.47 & 257.9 & 28.37 &  &  &  &  \\ 
24.3 & 1.5 & 39.2 & 4.3 & 805.9 & 88.6 & 18.7 & 2.06 &  &  &  &  &  &  \\ 
24.3 & 1.5 & 41.1 & 4.5 & 897.7 & 98.7 & 21.6 & 2.37 &  &  &  &  &  &  \\  
24.9 & 1.5 & 27.3 & 3.0 & 809.5 & 89.0 & 14.1 & 1.55 & 325.0 & 35.75 &  &  &  &  \\  
24.9 & 1.5 & 29.8 & 3.3 & 871.3 & 95.8 & 15.0 & 1.65 & 331.8 & 36.50 &  &  &  &  \\ 
25.1 & 1.5 & 23.2 & 2.6 &  &  & 12.5 & 1.38 &  &  &  &  &  &  \\  
25.4 & 1.4 & 22.9 & 2.5 &  &  & 11.7 & 1.28 &  &  &  &  &  &  \\  
25.5 & 1.4 & 28.0 & 3.1 & 923.7 & 101.6 & 14.5 & 1.59 & 312.2 & 34.35 &  &  &  &  \\  
26.0 & 1.4 & 19.8 & 2.2 &  &  & 9.5 & 1.04 & 364.9 & 40.14 &  &  &  &  \\  
26.1 & 1.4 & 20.3 & 2.2 &  &  & 10.5 & 1.16 & 316.1 & 34.77 &  &  &  &  \\  
26.5 & 1.4 & 22.1 & 2.4 &  &  & 10.0 & 1.10 & 345.2 & 37.97 &  &  &  &  \\ 
26.6 & 1.4 & 21.3 & 2.3 & 964.0 & 106.0 & 11.4 & 1.25 & 366.5 & 40.32 &  &  &  &  \\  
26.7 & 1.3 & 24.5 & 2.7 & 953.1 & 104.8 & 10.9 & 1.20 & 400.2 & 44.02 &  &  &  &  \\  
27.1 & 1.3 & 17.6 & 1.9 &  &  & 8.7 & 0.96 & 417.3 & 45.90 & 5.5 & 0.61 &  &  \\ 
27.2 & 1.3 & 19.0 & 2.1 & 889.3 & 97.8 & 8.9 & 0.98 &  &  & 1.8 & 0.20 &  &  \\ 
27.6 & 1.3 & 20.4 & 2.2 & 882.2 & 97.0 & 10.4 & 1.15 & 375.9 & 41.34 & 1.9 & 0.21 &  &  \\ 
27.6 & 1.3 & 18.0 & 2.0 & 972.9 & 107.0 & 12.4 & 1.36 & 406.3 & 44.70 & 2.2 & 0.25 &  &  \\  
27.7 & 1.3 & 18.1 & 2.0 & 953.8 & 104.9 & 10.4 & 1.15 & 427.3 & 47.00 & 1.6 & 0.18 &  &  \\  
27.9 & 1.3 &  &  & 972.0 & 106.9 & 12.8 & 1.41 & 408.4 & 44.93 & 2.8 & 0.30 &  &  \\ 
28.2 & 1.2 & 15.4 & 1.7 &  &  & 10.4 & 1.15 & 427.6 & 47.03 & 2.2 & 0.24 &  &  \\ 
28.2 & 1.2 & 16.3 & 1.8 &  &  & 11.3 & 1.24 &  &  & 5.5 & 0.60 &  &  \\  
28.7 & 1.2 & 16.7 & 1.8 & 931.7 & 102.5 & 17.8 & 1.96 &  &  & 8.5 & 0.94 &  &  \\ 
28.7 & 1.2 & 16.1 & 1.8 & 849.0 & 93.4 & 20.3 & 2.24 &  &  & 9.2 & 1.02 &  &  \\ 
28.7 & 1.2 & 15.5 & 1.7 &  &  & 17.5 & 1.92 & 441.7 & 48.59 & 9.0 & 0.99 &  &  \\ 
28.9 & 1.2 &  &  & 859.7 & 94.6 & 21.6 & 2.38 & 402.9 & 44.32 & 15.2 & 1.67 &  &  \\  
29.0 & 1.2 &  &  &  &  & 21.6 & 2.37 & 403.2 & 44.35 & 11.7 & 1.28 &  &  \\ 
29.2 & 1.2 & 13.7 & 1.5 &  &  & 21.2 & 2.33 & 411.6 & 45.27 & 11.0 & 1.21 &  &  \\ 
29.3 & 1.2 & 15.3 & 1.7 &  &  & 23.5 & 2.59 &  &  & 14.1 & 1.55 &  &  \\  
29.7 & 1.1 & 14.7 & 1.6 & 768.2 & 84.5 & 39.4 & 4.33 &  &  & 29.5 & 3.25 &  &  \\ 
29.7 & 1.1 & 15.1 & 1.7 &  &  & 40.4 & 4.45 & 382.4 & 42.06 & 29.6 & 3.26 &  &  \\ 
29.8 & 1.1 & 10.5 & 1.2 & 792.3 & 87.2 & 41.9 & 4.61 & 406.0 & 44.66 & 30.8 & 3.38 &  &  \\  
29.9 & 1.1 & 17.6 & 1.9 & 741.1 & 81.5 & 44.4 & 4.88 & 425.7 & 46.83 & 31.7 & 3.48 &  &  \\  
30.0 & 1.1 &  &  & 737.4 & 81.1 &  &  & 419.9 & 46.19 & 25.2 & 2.77 &  &  \\ 
30.2 & 1.1 & 16.5 & 1.8 & 720.6 & 79.3 & 54.0 & 5.94 & 393.0 & 43.23 & 44.7 & 4.92 &  &  \\ 
30.2 & 1.1 & 14.3 & 1.6 &  &  & 46.8 & 5.15 & 407.3 & 44.80 & 39.2 & 4.31 &  &  \\ 
30.3 & 1.1 & 13.8 & 1.5 &  &  & 48.0 & 5.28 &  &  & 37.0 & 4.06 &  &  \\ 
30.6 & 1.1 & 14.9 & 1.6 &  &  & 70.2 & 7.72 & 379.4 & 41.73 & 62.7 & 6.90 &  &  \\ 
30.7 & 1.1 & 18.0 & 2.0 & 684.7 & 75.3 & 74.1 & 8.15 &  &  & 65.4 & 7.19 &  &  \\ 
30.8 & 1.1 & 15.0 & 1.7 &  &  & 71.9 & 7.91 & 404.8 & 44.53 & 66.4 & 7.30 &  &  \\  
30.9 & 1.1 & 16.7 & 1.8 &  &  & 75.0 & 8.25 & 375.0 & 41.25 & 72.1 & 7.93 &  &  \\ 
31.0 & 1.1 & 16.5 & 1.8 & 704.3 & 77.5 & 85.4 & 9.39 & 369.2 & 40.61 & 71.8 & 7.90 &  &  \\  
31.6 & 1.0 & 16.0 & 1.8 & 499.1 & 54.9 & 108.1 & 11.89 &  &  & 116.7 & 12.83 &  &  \\  
31.7 & 1.0 & 15.4 & 1.7 & 468.9 & 51.6 & 108.8 & 11.97 & 323.1 & 35.54 & 114.9 & 12.63 &  &  \\ 
31.7 & 1.0 & 15.3 & 1.7 & 483.9 & 53.2 & 101.1 & 11.12 & 311.4 & 34.25 & 100.1 & 11.01 &  &  \\  
31.8 & 1.0 & 15.0 & 1.7 & 438.6 & 48.2 & 105.7 & 11.62 & 313.5 & 34.48 & 115.3 & 12.68 &  &  \\ 
31.9 & 1.0 & 14.1 & 1.6 &  &  & 110.0 & 12.10 & 299.6 & 32.96 & 106.4 & 11.70 &  &  \\ 
32.0 & 1.0 & 14.5 & 1.6 & 399.6 & 44.0 & 96.1 & 10.57 & 325.4 & 35.80 & 105.2 & 11.57 &  &  \\ 
32.6 & 0.9 & 14.2 & 1.6 & 396.6 & 43.6 & 139.3 & 15.33 & 273.5 & 30.08 & 153.3 & 16.86 &  &  \\ 
32.9 & 0.9 & 13.6 & 1.5 & 389.7 & 42.9 & 142.7 & 15.70 & 271.3 & 29.84 & 149.9 & 16.48 &  &  \\  
33.0 & 0.9 & 12.8 & 1.4 & 326.7 & 35.9 & 128.7 & 14.16 & 266.8 & 29.35 & 156.3 & 17.20 &  &  \\  
33.6 & 0.9 & 14.8 & 1.6 & 325.7 & 35.8 & 174.1 & 19.15 & 233.2 & 25.65 & 203.7 & 22.40 &  &  \\ 
33.8 & 0.9 & 14.5 & 1.6 & 316.4 & 34.8 & 179.3 & 19.72 & 224.8 & 24.73 & 198.3 & 21.81 &  &  \\  
33.9 & 0.8 & 12.9 & 1.4 & 299.7 & 33.0 & 152.7 & 16.79 & 216.9 & 23.85 & 193.5 & 21.28 &  &  \\ 
33.9 & 0.8 & 13.7 & 1.5 &  &  & 180.2 & 19.82 & 210.7 & 23.18 & 224.0 & 24.64 &  &  \\  
34.6 & 0.8 & 13.4 & 1.5 &  &  & 190.4 & 20.95 &  &  & 234.7 & 25.82 &  &  \\  
34.7 & 0.8 & 11.9 & 1.3 & 243.3 & 26.8 & 193.7 & 21.31 & 176.1 & 19.37 & 227.9 & 25.07 &  &  \\ 
34.8 & 0.8 & 14.6 & 1.6 & 237.6 & 26.1 & 189.8 & 20.88 & 166.9 & 18.36 & 251.0 & 27.61 &  &  \\ 
34.8 & 0.8 & 14.9 & 1.6 & 263.5 & 29.0 & 218.2 & 24.00 & 165.5 & 18.20 & 283.4 & 31.17 &  &  \\  
35.5 & 0.7 & 14.7 & 1.6 & 208.2 & 22.9 & 227.3 & 25.00 & 180.0 & 19.80 & 281.7 & 30.99 &  &  \\ 
35.6 & 0.7 & 10.4 & 1.2 & 206.4 & 22.7 & 209.3 & 23.02 & 155.8 & 17.14 & 249.7 & 27.46 &  &  \\  
35.7 & 0.7 & 13.4 & 1.5 & 232.3 & 25.6 & 192.9 & 21.22 & 137.7 & 15.15 & 260.6 & 28.66 &  &  \\ 
35.7 & 0.7 & 13.8 & 1.5 &  &  & 241.4 & 26.56 & 128.6 & 14.15 & 320.4 & 35.25 &  &  \\  
36.3 & 0.7 & 13.2 & 1.5 &  &  & 236.9 & 26.06 &  &  & 304.3 & 33.47 &  &  \\  
36.3 & 0.7 & 13.8 & 1.5 & 202.4 & 22.3 & 233.8 & 25.72 & 137.9 & 15.17 & 291.6 & 32.07 &  &  \\ 
36.4 & 0.7 & 12.9 & 1.4 & 193.2 & 21.3 & 222.5 & 24.47 & 136.6 & 15.03 & 283.1 & 31.14 &  &  \\ 
36.4 & 0.7 & 9.2 & 1.0 & 188.2 & 20.7 & 222.9 & 24.52 & 129.3 & 14.22 & 271.1 & 29.82 &  &  \\ 
36.6 & 0.7 & 12.9 & 1.4 & 185.0 & 20.4 & 204.6 & 22.50 & 126.8 & 13.95 & 273.8 & 30.12 &  &  \\  
36.6 & 0.7 & 12.5 & 1.4 &  &  & 240.5 & 26.45 & 118.6 & 13.04 & 309.8 & 34.08 & 2.2 & 0.2 \\ 
36.9 & 0.6 & 12.3 & 1.4 &  &  & 198.0 & 21.78 & 143.5 & 15.78 &  &  &  &  \\  
37.4 & 0.6 & 10.5 & 1.2 &  &  & 215.9 & 23.75 & 121.7 & 13.38 & 293.5 & 32.28 &  &  \\ 
37.4 & 0.6 & 11.1 & 1.2 & 149.8 & 16.5 & 227.6 & 25.04 & 148.6 & 16.35 & 301.7 & 33.19 & 2.5 & 0.3 \\  
37.8 & 0.6 & 12.7 & 1.4 & 162.6 & 17.9 & 216.0 & 23.76 & 154.3 & 16.98 &  &  &  &  \\  
38.3 & 0.5 & 11.7 & 1.3 & 154.3 & 17.0 & 252.4 & 27.76 & 137.6 & 15.14 & 343.3 & 37.76 & 12.9 & 1.4 \\  
38.3 & 0.5 & 13.4 & 1.5 & 167.0 & 18.4 & 261.6 & 28.77 & 225.1 & 24.76 & 357.8 & 39.35 & 9.8 & 1.1 \\ 
38.5 & 0.5 & 12.3 & 1.4 &  &  & 223.9 & 24.63 & 236.8 & 26.05 &  &  & 15.4 & 1.7 \\ 
39.1 & 0.5 & 11.6 & 1.3 & 149.3 & 16.4 & 244.2 & 26.86 & 223.8 & 24.62 & 328.5 & 36.13 & 22.6 & 2.5 \\  
39.1 & 0.5 & 11.4 & 1.3 & 131.1 & 14.4 & 218.4 & 24.02 & 295.7 & 32.53 & 304.2 & 33.46 & 20.5 & 2.3 \\  
39.3 & 0.5 & 10.5 & 1.2 & 128.9 & 14.2 & 206.3 & 22.69 & 259.5 & 28.54 &  &  & 30.1 & 3.3 \\  
39.9 & 0.4 & 12.5 & 1.4 & 127.9 & 14.1 & 198.7 & 21.86 & 275.0 & 30.25 & 294.7 & 32.42 & 34.2 & 3.8 \\  
39.9 & 0.4 & 10.6 & 1.2 &  &  &  &  & 321.4 & 35.35 & 290.4 & 31.95 & 41.5 & 4.6 2\\  
40.1 & 0.4 & 9.3 & 1.0 &  &  & 172.8 & 19.01 &  &  &  &  & 46.4 & 5.1 \\  
40.7 & 0.4 & 10.0 & 1.1 &  &  & 203.9 & 22.43 & 314.4 & 34.58 & 302.5 & 33.27 & 66.5 & 7.3 \\  
40.7 & 0.4 & 10.4 & 1.2 & 120.4 & 13.3 & 185.8 & 20.44 & 449.0 & 49.38 & 283.8 & 31.22 & 63.2 & 7.0 \\ 
40.9 & 0.4 & 8.6 & 0.9 & 98.9 & 10.9 & 160.1 & 17.61 & 436.9 & 48.05 &  &  & 70.6 & 7.8 \\  
41.5 & 0.3 & 11.0 & 1.2 & 111.1 & 12.2 & 180.5 & 19.85 &  &  & 274.6 & 30.20 & 91.5 & 10.1 \\ 
41.5 & 0.3 & 10.3 & 1.1 & 116.8 & 12.9 & 179.3 & 19.72 &  &  & 275.2 & 30.27 & 98.9 & 10.9 \\ 
41.7 & 0.3 & 9.4 & 1.0 & 105.3 & 11.6 & 157.8 & 17.35 & 515.5 & 41.24 &  &  & 93.8 & 10.3 \\

\hline
\end{tabular}
\end{center}
\end{table*}

\begin{table*}[t]
\tiny
\caption{Measured cross-sections of the ${}^{204m}$Pb,${}^{ 203}$Pb,${}^{ 202m}$Pb, ${}^{201}$Pb, and ${}^{200}$Pb reactions (ser. 2)}
\centering
\begin{center}
\begin{tabular}{|p{0.25in}|p{0.25in}|p{0.4in}|p{0.4in}|p{0.4in}|p{0.4in}|p{0.4in}|p{0.4in}|p{0.4in}|p{0.4in}|p{0.4in}|p{0.4in}|}
\hline 
\multicolumn{2}{|p{0.5in}|}{\textbf{E $\pm$  $\Delta $E(MeV)}} & \multicolumn{10}{|p{4.0in}|}{\textbf{Cross-sections ($\sigma$) $\pm$ $\Delta\sigma$ (mb)}}   \\ \hline 
\multicolumn{2}{|p{0.5in}|}{} & \multicolumn{2}{|p{0.8in}|}{${}^{204m}$Pb} & \multicolumn{2}{|p{0.8in}|}{${}^{203}$Pb} & \multicolumn{2}{|p{0.8in}|}{${}^{202m}$Pb} & \multicolumn{2}{|p{0.8in}|}{${}^{201}$Pb} & \multicolumn{2}{|p{0.8in}|}{${}^{200}$Pb}   \\ \hline 
36.4 & 0.3 & 11.4 & 1.4 & 204.2 & 23.0 & 227.2 & 24.6 & 113.1 & 12.7 & 270.7 & 29.3   \\  
35.2 & 0.3 & 11.5 & 1.3 & 232.6 & 26.0 & 197.7 & 21.4 & 130.8 & 14.3 & 251.0 & 27.2   \\ 
34.0 & 0.3 & 13.6 & 1.5 & 275.8 & 30.5 & 176.2 & 19.1 & 166.6 & 18.1 & 206.1 & 22.3   \\  
32.8 & 0.3 & 12.6 & 1.4 & 362.0 & 40.3 & 137.3 & 14.9 & 217.5 & 23.6 & 211.1 & 22.9   \\ 
31.5 & 0.4 & 14.3 & 1.6 & 468.1 & 51.0 & 103.6 & 11.2 & 275.5 & 29.9 & 88.5 & 9.6   \\ 
30.2 & 0.4 & 14.7 & 1.6 & 621.7 & 67.8 & 59.3 & 6.5 & 316.1 & 34.3 & 44.1 & 4.8   \\ 
28.8 & 0.4 & 16.8 & 1.9 & 769.6 & 83.7 & 26.4 & 2.9 & 355.0 & 38.5 & 11.7 & 1.4   \\  
27.4 & 0.5 & 18.8 & 2.6 & 836.4 & 90.9 & 10.2 & 1.2 & 345.7 & 37.6 & 0.5 & 0.5   \\ 
25.9 & 0.5 & 22.5 & 2.7 & 858.9 & 93.4 & 11.8 & 1.4 & 338.7 & 37.0 &  &    \\  
24.4 & 0.6 & 28.0 & 3.1 & 768.7 & 83.5 & 17.0 & 1.9 & 287.3 & 31.3 &  &    \\  
22.8 & 0.6 & 45.7 & 5.2 & 686.2 & 74.7 & 26.6 & 2.9 & 223.7 & 24.3 &  &  \\  
21.1 & 0.7 & 73.7 & 8.2 & 532.2 & 58.0 & 31.0 & 3.4 & 127.8 & 14.0 &  &    \\  
19.3 & 0.8 & 96.2 & 10.8 & 244.2 & 26.8 & 28.6 & 3.2 & 22.8 & 3.5 &  &    \\  
17.8 & 0.9 & 75.8 & 8.2 & 65.1 & 8.2 & 25.7 & 2.8 &  &  &  &    \\  
16.2 & 0.9 & 63.4 & 6.9 & 11.1 & 3.4 & 12.8 & 1.4 &  &  &  &    \\ 
14.6 & 1.0 & 33.4 & 3.6 & 20.3 & 2.9 & 10.4 & 1.2 &  &  &  &    \\ 
12.7 & 1.1 & 11.8 & 1.5 & 36.5 & 4.2 &  &  &  &  &  &    \\ 
10.6 & 1.3 & 0.6 & 0.3 & 40.2 & 4.5 &  &  &  &  &  &    \\ 
9.3 & 1.4 &  &  & 27.8 & 3.3 &  &  &  &  &  &    \\  
7.9 & 1.5 &  &  & 10.1 & 1.1 &  &  &  &  &  &    \\
\hline
\end{tabular}
\end{center}
\end{table*}

\begin{table*}[t]
\tiny
\caption{Measured cross-sections of the ${}^{202}$Tl, ${}^{201}$Tl, ${}^{200}$Tl, and ${}^{203}$Hg reactions (ser.2)}
\centering
\begin{center}
\begin{tabular}{|p{0.25in}|p{0.25in}|p{0.4in}|p{0.4in}|p{0.4in}|p{0.4in}|p{0.4in}|p{0.4in}|p{0.4in}|p{0.4in}|}
\hline 
\multicolumn{2}{|p{0.5in}|}{\textbf{E $\pm$  $\Delta $E(MeV)}} & \multicolumn{8}{|p{3.2in}|}{\textbf{Cross-sections ($\sigma$) $\pm$ $\Delta\sigma$ (mb)}}   \\ \hline 
\multicolumn{2}{|p{0.5in}|}{} & \multicolumn{2}{|p{0.8in}|}{${}^{202}$Tl} & \multicolumn{2}{|p{0.8in}|}{${}^{201}$Tl} & \multicolumn{2}{|p{0.8in}|}{${}^{200}$Tl} & \multicolumn{2}{|p{0.8in}|}{${}^{203}$Hg}   \\ \hline 
36.4 & 0.3 & 80.1 & 8.9 & 54.8 & 23.8 & 5.1 & 0.6 & 0.36 & 0.05   \\ 
35.2 & 0.3 & 77.8 & 8.7 &  &  & 2.5 & 0.3 & 0.26 & 0.07   \\  
34.0 & 0.3 & 67.6 & 7.6 & 51.8 & 28.4 & 2.9 & 0.3 & 0.25 & 0.07   \\  
32.8 & 0.3 & 64.8 & 7.3 &  &  & 1.5 & 0.2 & 0.13 & 0.07   \\ 
31.5 & 0.4 & 58.4 & 6.4 & 85.3 & 27.1 & 1.5 & 0.2 & 0.07 & 0.06   \\ 
30.2 & 0.4 & 46.6 & 5.4 & 79.9 & 27.0 & 0.6 & 0.1 &  &    \\ 
28.8 & 0.4 & 46.0 & 5.3 & 99.6 & 29.9 &  &  &  &    \\  
27.4 & 0.5 & 41.7 & 4.8 & 149.5 & 28.9 &  &  &  &    \\ 
25.9 & 0.5 & 33.7 & 4.0 & 98.9 & 23.9 &  &  &  &    \\  
24.4 & 0.6 & 30.8 & 3.6 & 72.0 & 23.9 &  &  &  &    \\  
22.8 & 0.6 & 26.4 & 3.2 & 112.8 & 29.2 &  &  &  &    \\ 
21.1 & 0.7 & 20.5 & 2.6 & 38.6 & 11.8 &  &  &  &    \\ 
19.3 & 0.8 & 14.6 & 1.8 & 1.0 & 8.7 &  &  &  &    \\ 
17.8 & 0.9 & 10.3 & 1.3 &  &  &  &  &  &    \\  
16.2 & 0.9 & 5.4 & 0.8 &  &  &  &  &  &    \\ 
14.6 & 1.0 & 2.2 & 0.3 &  &  &  &  &  &    \\  
12.7 & 1.1 & 0.8 & 0.2 &  &  &  &  &  &    \\  
10.6 & 1.3 &  &  &  &  &  &  &  &    \\ 
9.3 & 1.4 & 0.3 & 0.1 &  &  &  &  &  &    \\  
7.9 & 1.5 & 0.4 & 0.1 &  &  &  &  &  &    \\  

\hline
\end{tabular}
\end{center}
\end{table*}

\subsection{Integral yields}
\label{4.2}

From fits to our experimental excitation functions thick target physical yields \citep{9} were calculated (no saturation effects, obtained in an irradiation time that is very short compared to the half-life of the radionuclide). The integral yields for $^{201}Pb$, $^{202m}Pb$, $^{200}Pb$, $^{203}Pb$ and $^{204m}Pb$ (related to production of $^{201}Tl$ and $^{203}Pb$ ) are shown in Figs. 14-18 in comparison with the directly measured data in the literature. Only very few experimental thick target yield data were found in the literature.  The data reported for enriched and natural targets were normalized according to the real isotopic composition and to the contributing reactions.  Our calculated integral yield values for the production of $^{203}Pb$ and are higher than the direct yield data.  In cases of $^{202m}Pb$ and $^{201}Pb$ our calculated data are surprisingly lower. No experimental thick target yield data were found for $^{200}Pb$ in the literature. 

\begin{figure}[h]
\includegraphics[scale=0.3]{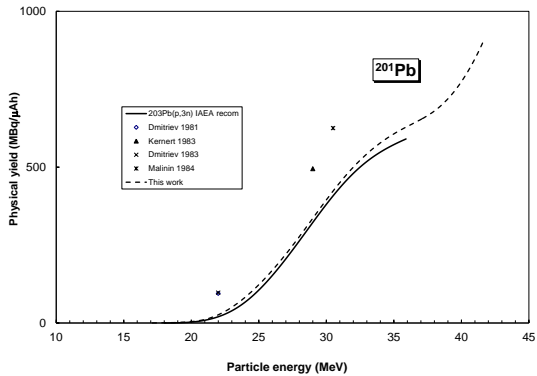}
\caption{Integral yields of $^{201}Pb$ calculated from the measured excitation functions in comparison with experimental thick target yields from literature}
\end{figure}

\begin{figure}[h]
\includegraphics[scale=0.3]{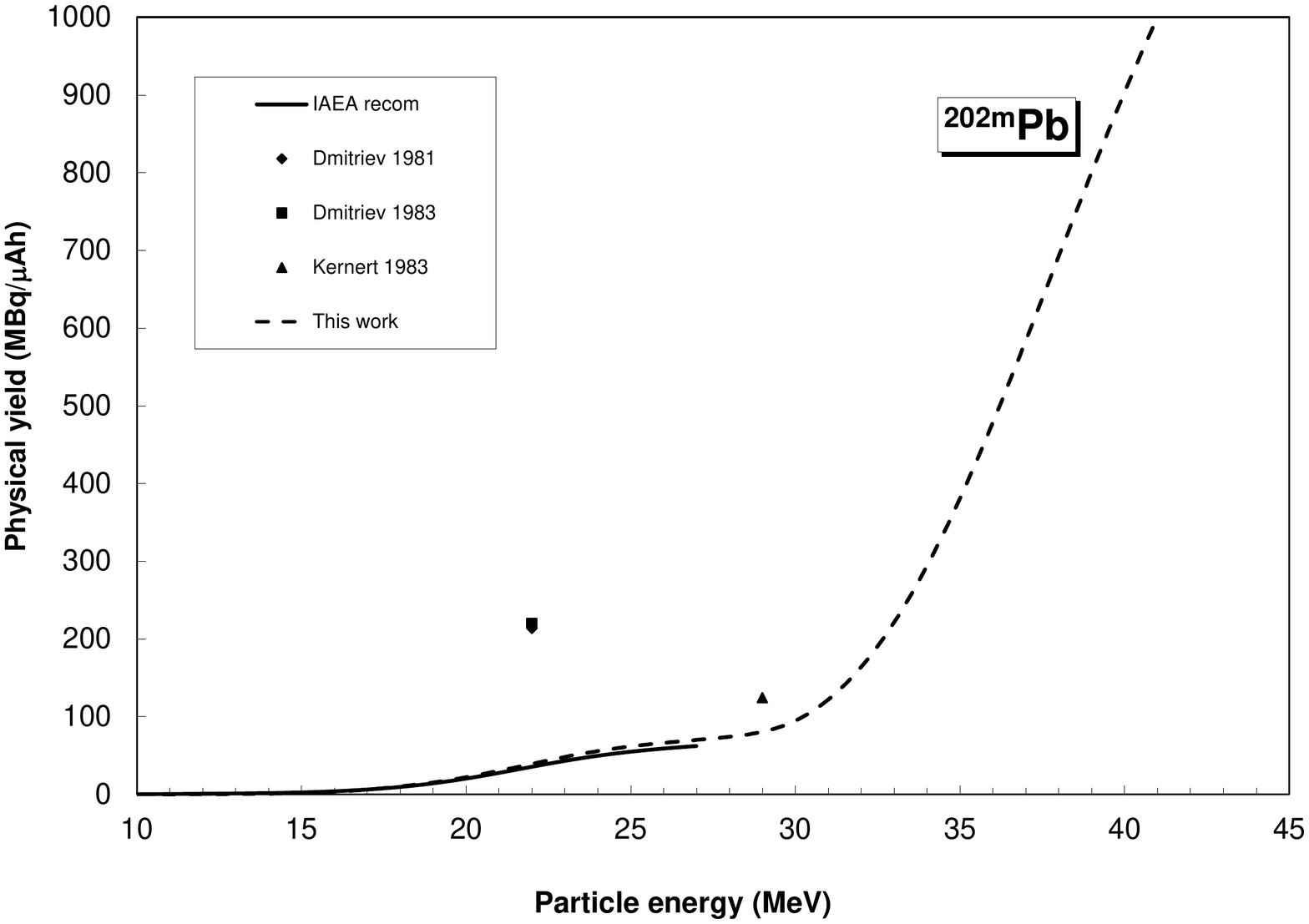}
\caption{Integral yields of $^{202m}Pb$ calculated from the measured excitation functions in comparison with experimental thick target yields from literature}
\end{figure}

\begin{figure}[h]
\includegraphics[scale=0.3]{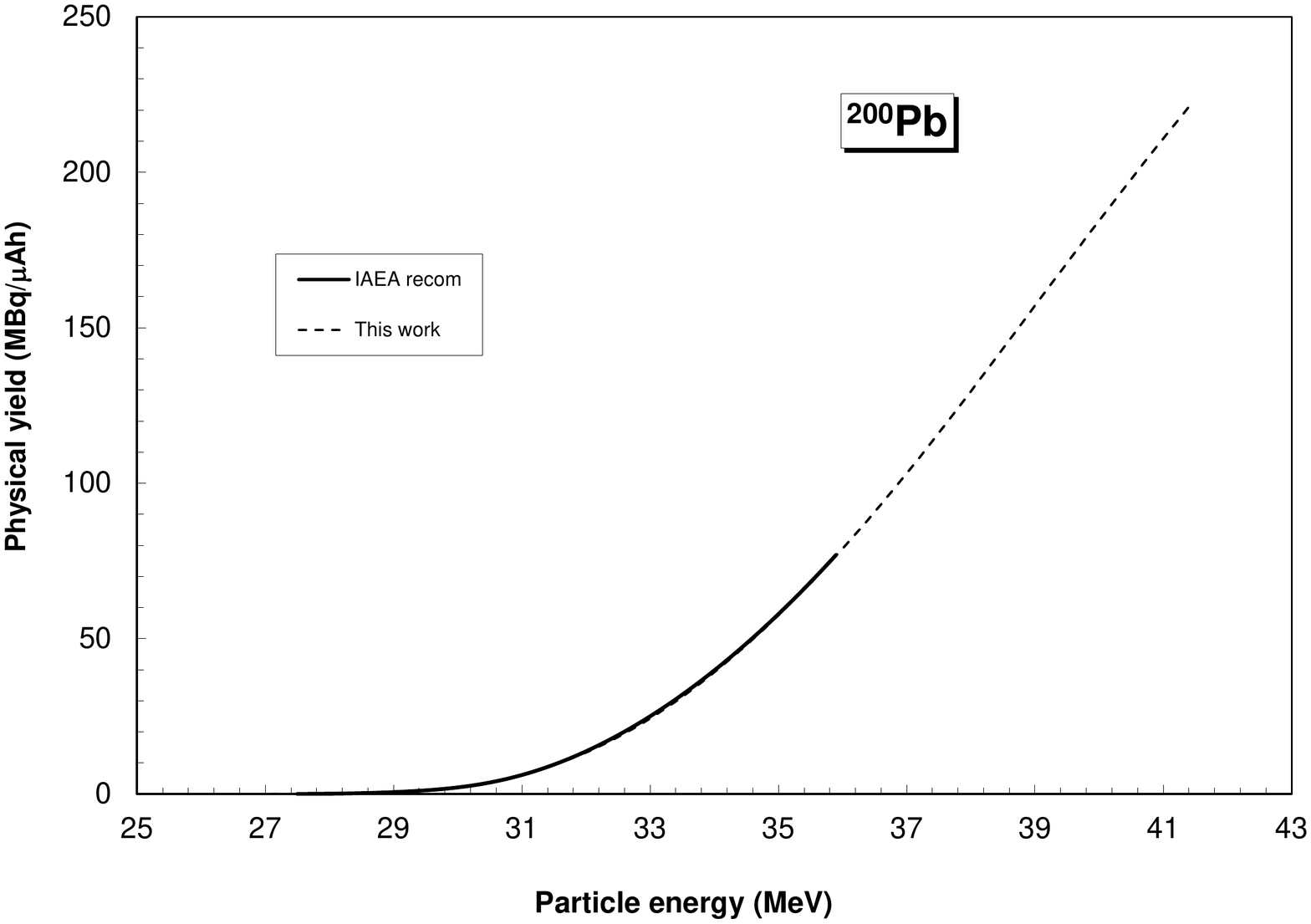}
\caption{Integral yields of $^{200}Pb$ calculated from the measured excitation functions in comparison with experimental thick target yields from literature}
\end{figure}

\begin{figure}[h]
\includegraphics[scale=0.3]{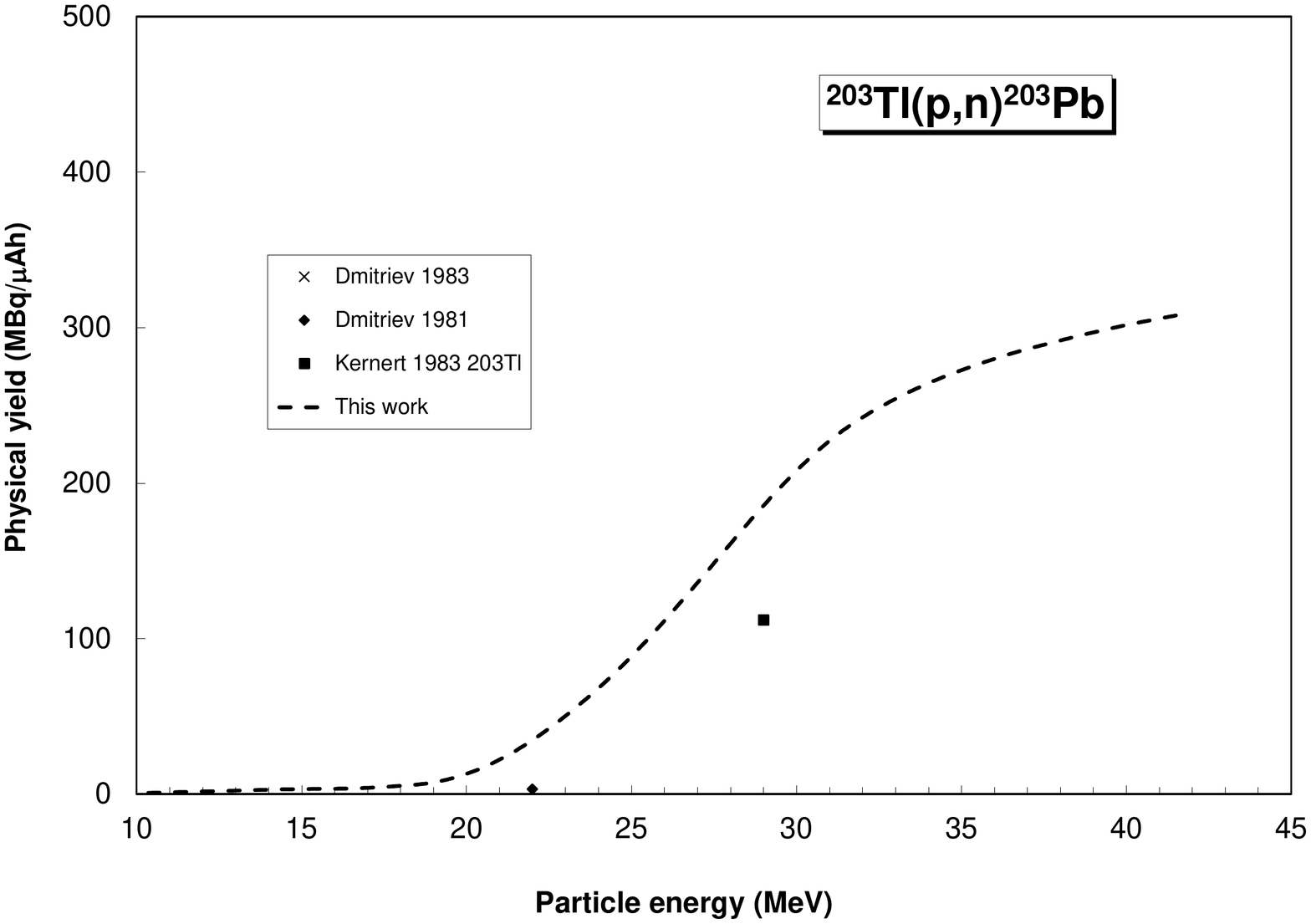}
\caption{Integral yields of $^{203}Pb$ calculated from the measured excitation functions in comparison with experimental thick target yields from literature}
\end{figure}

\begin{figure}[h]
\includegraphics[scale=0.3]{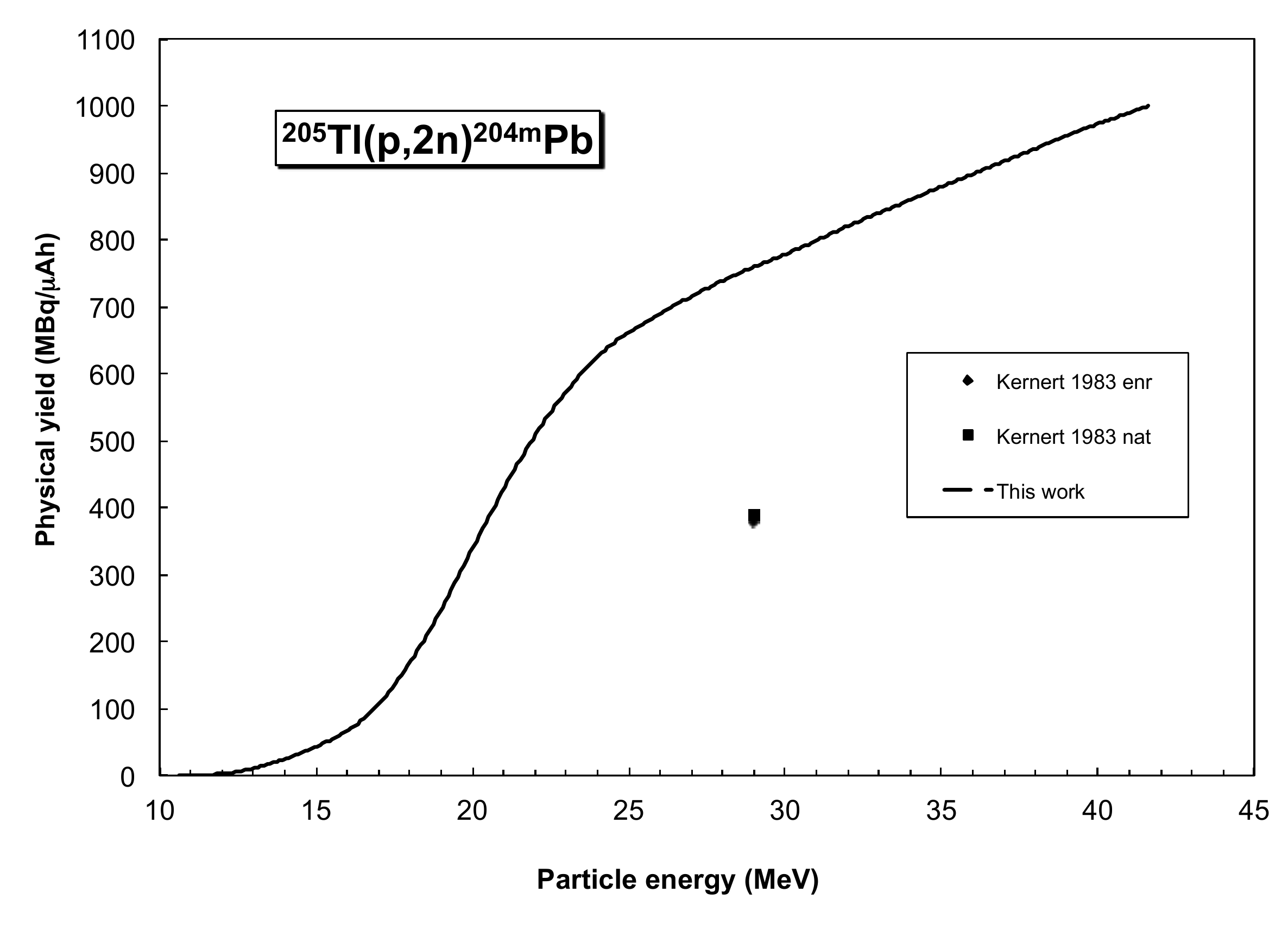}
\caption{Integral yields of $^{204m}Pb$ calculated from the measured excitation functions in comparison with experimental thick target yields from literature}
\end{figure}

\section{Summary and conclusions}
\label{5}

The principal aim of this investigation was re-measurement and evaluation of the cross-section data of the production of $^{201}Tl$. 
\begin{enumerate}
\item We present new excitation curves for proton induced reactions on $^{nat}Tl$ targets up to 42 MeV leading to the production of $^{204m}Pb$, $^{203}Pb$, $^{202m}Pb$, $^{201}Pb$, $^{200}Pb$, $^{199}Pb$, $^{202}Tl$ (dir, cum), $^{201}Tl$ (dir, cum), $^{200}Tl$(dir) and $^{203}Hg$.
\item Detailed literature search was done and the literature data were critically analyzed (the earlier errors were corrected, the derived data were changed to original experimental data).
\item	To upgrade the recommended cross-section data  new experimental data were involved (Hanna \citep{26,4}, this work).
\item To check the predictivity of the theoretical codes, new theoretical calculations were done based on new versions of the model codes. 
\end{enumerate}

\subsection{The new experimental data:}
\label{5.1}

As our excitation functions are correlated to monitor reactions re-measured over the whole energy range simultaneously the uncertainties in incident energy, average energy in each foil and number of bombarding particles are diminished.

\subsection{The evaluation process:}
\label{5.2}
\begin{enumerate}
\item Except a few old works the experimental data shows surprisingly good agreement.
\item The description of the theoretical codes is acceptable good.
\item The new recommended data from point of view of production of $^{201}Tl$ and the $^{200}Tl$ and $^{202}Tl$ impurities didn't differ significantly from recommended data in the IAEA medical library.
\item No proper thick target yield data were measured for validation of the excitation functions (low beam intensity measurements at around 30 MeV).
\item There are good review works discussing the optimization of the $^{201}Tl$ production and the chemical processes \citep{3,4,7,20,24,25,26,41,42,44,52,55,56}(. 
\item Except excitation functions of Lagunas Solar \citep{42}  measured up to 60 MeV  for  production of $^{200-204}Pb$ , no experimental data exists above 40 MeV 
\end{enumerate}

\subsection{Application of the new experimental and evaluated data:}
\label{5.3}
\begin{enumerate}
\item Optimization of the production of the medically used $^{201}Tl$.
\item The measured excitation functions can be used for determination of thallium in different materials by proton activation analysis \citep{15,63}.
\item To estimate the radiation dose caused by proton activation of the materials with thallium content (glasses, semiconductors, particle detectors, high temperature superconductors, etc.)
\item Application of $^{203}Pb$ emitting imaginable $\gamma$-ray (279 keV) for bio-distribution and targeting studies of the cytotoxic alpha-emitting $^{212}Pb$-radiolabeled compounds and antibodies \citep{13,21} . Preparation of $^{203}Pb$ compounds for studies on pathway and effects of lead pollution \citep{22} .
\item To upgrade the nuclear reaction model codes and their input parameters.
\end{enumerate}

\section{Acknowledgements}

This study was performed in the frame of the MTA-FWO (Vlaanderen) collaboration program. The authors thank the different research projects and their respective institutions for the practical help and providing the use of the facilities for this study. Special thanks to V. Semkova (IAEA) for valuable help in literature search.

\clearpage
\bibliographystyle{elsarticle-harv}
\bibliography{Tl+p}

\begin{thebibliography}{63}
\expandafter\ifx\csname natexlab\endcsname\relax\def\natexlab#1{#1}\fi
\expandafter\ifx\csname url\endcsname\relax
  \def\url#1{\texttt{#1}}\fi
\expandafter\ifx\csname urlprefix\endcsname\relax\def\urlprefix{URL }\fi

\bibitem[{Adam-Rebeles et~al.(2011)Adam-Rebeles, Hermanne, Van~den Winkel,
  T\'ark\'anyi, and Tak\'acs}]{1}
Adam-Rebeles, R., Hermanne, A., Van~den Winkel, P., T\'ark\'anyi, F., Tak\'acs,
  S., 2011. Activation cross section of deuteron induced reactions on natural
  thallium for the production of 203pb. Journal of the Korean Physical Society
  59, 1975--1978.

\bibitem[{Adam-Rebeles et~al.(2012)Adam-Rebeles, Van~den Winkel, Hermanne,
  T\'ark\'anyi, and Tak\'acs}]{2}
Adam-Rebeles, R., Van~den Winkel, P., Hermanne, A., T\'ark\'anyi, F., Tak\'acs,
  S., 2012. Experimental excitation functions of deuteron induced reactions
  onnatural thallium up to 50 mev. Nuclear Instruments \& Methods in Physics
  Research Section B (accepted).

\bibitem[{Afarideh et~al.(2004)Afarideh, Al-Jammaz, Arzumanov, Casale, Dudu,
  Haji-Saeid, Narasimhan, Schlyer, Solin, Takács, Van~den Winkel, Vera~Ruiz,
  and Zhou}]{3}
Afarideh, H., Al-Jammaz, I., Arzumanov, A., Casale, G., Dudu, D., Haji-Saeid,
  M., Narasimhan, D. V.~S., Schlyer, D., Solin, L., Takács, S., Van~den
  Winkel, P.~N., Vera~Ruiz, H., Zhou, W., 2004. Standardized high current solid
  targets dor cyclotron production of diagnostic and therapeutic radionuclides.
  Tech. rep., IAEA.

\bibitem[{Al-Saleh et~al.(2007)Al-Saleh, Al-Harbi, and Azzam}]{4}
Al-Saleh, F.~S., Al-Harbi, A.~A., Azzam, A., 2007. Yield and excitation
  function measurements of some nuclear reactions on natural thallium induced
  by protons leading to the production of medical radioisotopes 201tl and
  203pb. Radiochimica Acta 95~(3), 127--132.

\bibitem[{Andersen and Ziegler(1977)}]{5}
Andersen, H.~H., Ziegler, J.~F., 1977. Hydrogen stopping powers and ranges in
  all elements. The Stopping and ranges of ions in matter, Volume 3. The
  Stopping and ranges of ions in matter. Pergamon Press, New York.

\bibitem[{Bell and Skarsgard(1956)}]{6}
Bell, E., Skarsgard, H.~M., 1956. Cross sections of (p, xn) reactions in the
  isotope of lead and bismuth. Canadian Journal of Physics 34~(8), 745--766.

\bibitem[{Birattari et~al.(1982)Birattari, Bonardi, and Salomone}]{7}
Birattari, C., Bonardi, M., Salomone, A., 1982. Tl-201 production studies by
  tl-203 (p,3n) pb-201 and hg-202 (p,2n) nuclear-reactions. Journal of Labelled
  Compounds \& Radiopharmaceuticals 19~(11-1), 1330--1332.

\bibitem[{Blue et~al.(1978)Blue, Liu, and Smathers}]{8}
Blue, J.~W., Liu, D.~C., Smathers, J.~B., 1978. Thallium-201 production with
  the idle beam from neutron therapy. Medical Physics 5~(6), 532--5.

\bibitem[{Bonardi(1987)}]{9}
Bonardi, M., 1987. The contribution to nuclear data for biomedical radioisotope
  production from the milan cyclotron facility.

\bibitem[{Bonardi et~al.(1982)Bonardi, Birattari, and Salomone}]{10}
Bonardi, M., Birattari, C., Salomone, A., 1982. 201-tl production for medical
  use by (p,xn) nuclear reactions on tl and hg natural and enriched targets.
  In: Conf.on Nucl.Data for Sci.and Technol. p. 916.

\bibitem[{Browne and Firestone(1986)}]{11}
Browne, E., Firestone, R.~B., 1986. Table of Radioactive Isotopes. John Wiley
  \& Sons, New York.

\bibitem[{Canderias-Cruz and Okamoto(1987)}]{12}
Canderias-Cruz, D., Okamoto, J., 1987. Nuclear data for medical radioisotopes
  produced by accelerators-status and compilation. In: IAEA Consultants’
  Meeting on Data Requirements for Medical Radioisotope Production. pp.
  INDC(NDS)--195--GZ, 1998, p.145.

\bibitem[{Chappell et~al.(2000)Chappell, Dadachova, Milenic, Garmestani, Wu,
  and Brechbiel}]{13}
Chappell, L.~L., Dadachova, E., Milenic, D.~E., Garmestani, K., Wu, C.~C.,
  Brechbiel, M.~W., 2000. Synthesis, characterization, and evaluation of a
  novel bifunctional chelating agent for the lead isotopes pb-203 and pb-212.
  Nuclear Medicine and Biology 27~(1), 93--100.

\bibitem[{Comar and Crouzel(1975)}]{14}
Comar, D., Crouzel, C., 1975. Preparation of carrier-free radioactive thallium
  for medical use. Radiochemical and Radioanalytical Letters 23~(3), 131--137.

\bibitem[{De~Brucker et~al.(1989)De~Brucker, Dewaele, Strijckmans, and
  Vandecasteele}]{15}
De~Brucker, N., Dewaele, J., Strijckmans, K., Vandecasteele, C., 1989.
  Determination of thallium in zinc by proton activation-analysis. Analytica
  Chimica Acta 220~(1), 93--102.

\bibitem[{Dityuk et~al.(1998)Dityuk, Konobeyev, Lunev, and Shubin}]{16}
Dityuk, A.~I., Konobeyev, A.~Y., Lunev, V.~P., Shubin, Y.~N., 1998. New version
  of the advanced computer code alice-ippe. Tech. rep., IAEA.

\bibitem[{Dmitriev et~al.(1976)Dmitriev, Molin, Dmitrieva, and Panarin}]{17}
Dmitriev, P.~P., Molin, G.~A., Dmitrieva, Z.~P., Panarin, M.~V., 1976. Yields
  of tl-200, tl-201, tl-202, and tl-204 during proton and deuteron irradiation
  of mercury. Soviet Atomic Energy 41~(6), 1091--1093.

\bibitem[{Dmitriev and Zaitseva(1996)}]{18}
Dmitriev, S.~N., Zaitseva, N.~G., 1996. Radionuclides for biomedical studies.
  nuclear data and production methods in charged-particle accelerators. Physics
  of Particles and Nuclei 27~(4), 403--427.

\bibitem[{Erdtmann and Soyka(1979)}]{19}
Erdtmann, G., Soyka, W., 1979. The Gamma Rays of the Radionuclides. Verlag
  Chemie, Weinheim, New-York.

\bibitem[{Fernandes and da~Silva(1992)}]{20}
Fernandes, L., da~Silva, C. P.~G., 1992. A study of irradiation conditions of
  mercury target with protons to obtain thallium-201. Journal of Labelled
  Compounds and Radiopharmaceuticals 31~(12), 967--971.

\bibitem[{Garmestania et~al.(2005)Garmestania, Milenica, Bradya, Plascjakb, and
  Brechbiela}]{21}
Garmestania, K., Milenica, D.~E., Bradya, E.~D., Plascjakb, P.~S., Brechbiela,
  M.~W., 2005. Purification of cyclotron-produced 203pb for labeling herceptin.
  Nuclear Medicine and Biology 32~(3), 301--305.

\bibitem[{Girardi et~al.(1975)Girardi, Goetz, Sabbioni, Marafante, Merlini,
  Acerbi, Birattari, Castiglioni, and Resmini}]{22}
Girardi, F., Goetz, L., Sabbioni, E., Marafante, E., Merlini, M., Acerbi, E.,
  Birattari, C., Castiglioni, M., Resmini, F., 1975. Preparation of pb-203
  compounds for studies on pathway and effects of lead pollution. International
  Journal of Applied Radiation and Isotopes 26~(5), 267--277.

\bibitem[{Gloris et~al.(2001)Gloris, Michel, Sudbrock, Herpers, Malmborg, and
  Holmqvist}]{23}
Gloris, M., Michel, R., Sudbrock, F., Herpers, U., Malmborg, P., Holmqvist, B.,
  2001. Proton-induced production of residual radionuclides in lead at
  intermediate energies. Nuclear Instruments \& Methods in Physics Research
  Section a-Accelerators Spectrometers Detectors and Associated Equipment
  463~(3), 593--633.

\bibitem[{Groppi et~al.(2001)Groppi, Bonardi, Birattari, Gini, Severgnini, and
  Mainardi}]{24}
Groppi, F., Bonardi, M., Birattari, C., Gini, L., Severgnini, M., Mainardi, C.,
  2001. A rapid improved method for gamma-spectrometry determination of
  thallium-202 impurities in [thallium-201] labeled radiopharmaceuticals. Tech.
  rep., Istituto Nazionale di Fisica Nucleare.

\bibitem[{Haji-Saeid et~al.(2009)Haji-Saeid, Pilai, Ruth, Schleyer, Van~den
  Winkel, Vora, Capote-Noy, Carroll, Clark, Comor, Dehnel, Ferrieri, Finn,
  Fowler, Schueller, and Tárkányi}]{25}
Haji-Saeid, M., Pilai, M. R.~A., Ruth, T.~J., Schleyer, D.~J., Van~den Winkel,
  P., Vora, M.~M., Capote-Noy, R., Carroll, L., Clark, J.~C., Comor, J.,
  Dehnel, M., Ferrieri, R., Finn, R.~D., Fowler, J.~S., Schueller, M.~J.,
  Tárkányi, F., 2009. Cyclotron produced radionuclieds: Physical
  characteristics and production methods. Tech. rep., IAEA.

\bibitem[{Hanna et~al.(1977)Hanna, Leigh, and Burch}]{26}
Hanna, R.~W., Leigh, J.~R., Burch, W.~M., 1977. Production and separation of
  tl-201 suitable for clinical myocardial imaging. Australasian Radiology
  21~(4), 387--393.

\bibitem[{Herman et~al.(2007)Herman, Capote, Carlson, Oblozinsky, Sin, Trkov,
  Wienke, and Zerkin}]{27}
Herman, M., Capote, R., Carlson, B.~V., Oblozinsky, P., Sin, M., Trkov, A.,
  Wienke, H., Zerkin, V., 2007. Empire: Nuclear reaction model code system for
  data evaluation. Nuclear Data Sheets 108~(12), 2655--2715.

\bibitem[{Hermanne et~al.(2001)Hermanne, Gul, Mustafa, Nortier, Oblozinsky,
  Qaim, Scholten, Shubin, T\'ark'anyi, Tak'acs, and Youxiang}]{28}
Hermanne, A., Gul, K., Mustafa, M.~G., Nortier, M., Oblozinsky, P., Qaim,
  S.~M., Scholten, B., Shubin, Y.~N., T\'ark'anyi, F., Tak'acs, S., Youxiang,
  Z., 2001. Production cross-sections for diagnostic radioisotopes (chapter 5).
  gamma emitters (5.1). Tech. rep., IAEA.

\bibitem[{Hermanne et~al.(1992)Hermanne, Walravens, and Cichelli}]{29}
Hermanne, A., Walravens, N., Cichelli, O., 1992. Optimisation of isotope
  production by cross section determination. In: Qaim, S.~M. (Ed.),
  International conference on nuclear data for science and technology.
  Springer, pp. 616--618.

\bibitem[{Iljinov et~al.(1993)Iljinov, Semenov, Semenova, Sobolevsky, and
  Udovenko}]{30}
Iljinov, A.~S., Semenov, V.~G., Semenova, M.~P., Sobolevsky, N.~M., Udovenko,
  L.~V., 1993. Production of Radionuclides at Intermediate Energies
  Interactions of Protons with Targets from I to Am. Vol.~13. Springer,
  Berlin-Heidelberg-New York.

\bibitem[{Janni(1966)}]{32}
Janni, J.~F., 1966. Calculations of energy loss, range, path length,
  straggling, multiple scattering and the probability of inelastic nuclear
  collisions for 0.1 to 1000 mev protons. Tech. rep., Air Force Weapons
  Laboratory.

\bibitem[{Kaplan et~al.(2009)Kaplan, Aydin, Tel, and Sarer}]{33}
Kaplan, A., Aydin, A., Tel, E., Sarer, B., 2009. Equilibrium and
  pre-equilibrium emissions in proton-induced reactions on tl-203,tl-205.
  Pramana-Journal of Physics 72~(2), 343--353.

\bibitem[{Kernert et~al.(1983)Kernert, Peters, and Schweickert}]{34}
Kernert, N., Peters, J.~W., Schweickert, H., 1983. Test production of 201pb
  with protons. Jul-Spez 202, 119.

\bibitem[{Koning et~al.(2007)Koning, Hilaire, and Duijvestijn}]{35}
Koning, A.~J., Hilaire, S., Duijvestijn, M.~C., 2007. Talys-1.0.

\bibitem[{Koning and Rochman(2011)}]{36}
Koning, A.~J., Rochman, D., 2011. Talys-based evaluated nuclear data library
  version 4.

\bibitem[{Kuhnhenn et~al.(2001)Kuhnhenn, Herpers, Glasser, Michel, Kubik, and
  Suter}]{37}
Kuhnhenn, J., Herpers, U., Glasser, W., Michel, R., Kubik, P.~W., Suter, M.,
  2001. Thin target cross sections for proton-induced formation of
  radionuclides from lead for e(p) <= 71 mev. Radiochimica Acta 89~(11-12),
  697--702.

\bibitem[{Kurenkov et~al.(1995)Kurenkov, Lunev, Masterov, and Shubin}]{38}
Kurenkov, N.~V., Lunev, V.~P., Masterov, V.~S., Shubin, Y.~N., 1995.
  Excitation-functions for the formation of the neutron-deficient nuclei
  tl-201, pb-201 and bi-201 - calculated and experimental-data. Applied
  Radiation and Isotopes 46~(1), 29--37.

\bibitem[{Lagunas-Solar et~al.(1980)Lagunas-Solar, Jungerman, and Paulson}]{40}
Lagunas-Solar, M.~C., Jungerman, J.~A., Paulson, D.~W., 1980. Tl-201 yields and
  excitation-functions for the lead radioactivities produced by irradiation of
  tl-205 with 34-60 mev protons. International Journal of Applied Radiation and
  Isotopes 31~(2), 117--121.

\bibitem[{Lagunas-Solar et~al.(1981)Lagunas-Solar, Little, and Jungerman}]{41}
Lagunas-Solar, M.~C., Little, F.~E., Jungerman, J.~A., 1981. Proton-induced
  reactions on natural pb targets - a potential new cyclotron method for t1-201
  production. International Journal of Applied Radiation and Isotopes 32~(11),
  817--822.

\bibitem[{Lagunas-Solar et~al.(1978)Lagunas-Solar, Little, Jungerman, Peak, and
  Theus}]{42}
Lagunas-Solar, M.~C., Little, F.~E., Jungerman, J.~A., Peak, N.~T., Theus,
  R.~M., 1978. Thallium-201 yields and excitation functions for the lead
  induced radioactivities produced by irradiation of natural thallium with
  15-60 mev protons. Int. J. Appl. Radiat. Isot. 29, 159--165.

\bibitem[{Lebowitz et~al.(1975)Lebowitz, Greene, Fairchild, Bradleymoore,
  Atkins, Ansari, Richards, and Belgrave}]{43}
Lebowitz, E., Greene, M.~W., Fairchild, R., Bradleymoore, P.~R., Atkins, H.~L.,
  Ansari, A.~N., Richards, P., Belgrave, E., 1975. Tl-201 for medical use .1.
  Journal of Nuclear Medicine 16~(2), 151--155.

\bibitem[{Malinin et~al.(1984)Malinin, Kozlova, Sevastyanova, Kharlamov,
  Chursin, Kochetkov, Gladun, Chumikov, Krasnov, Konyakhin, Kulygin, and
  Abdukayumov}]{44}
Malinin, A.~B., Kozlova, M.~D., Sevastyanova, A.~S., Kharlamov, V.~T., Chursin,
  G.~P., Kochetkov, V.~L., Gladun, V.~T., Chumikov, G.~N., Krasnov, N.~N.,
  Konyakhin, N.~A., Kulygin, V.~M., Abdukayumov, M.~A., 1984. Production of
  no-carrier-added tl-201. International Journal of Applied Radiation and
  Isotopes 35~(7), 685--687.

\bibitem[{Mihailescu et~al.(2007)Mihailescu, Nechifor, Straticiuc, and
  Bercea}]{45}
Mihailescu, D., Nechifor, C.-D., Straticiuc, M., Bercea, M., 2007. Cross
  sections and protons optimum energy ranges for some medical radioisotopes
  production. Tech. rep., STIINŢIFICE ALE UNIVERSITĂŢII “AL. I. CUZA”
  IAŞI.

\bibitem[{Milazzo-Colli et~al.(1975)Milazzo-Colli, Braga-Marcazzan, and
  Milazzo}]{46}
Milazzo-Colli, L., Braga-Marcazzan, G.~M., Milazzo, M., 1975. Further
  measurements of probability of alpha-cluster pre-formation by means of (p,
  alpha) reactions in heavy elements. Nuovo Cimento Della Societa Italiana Di
  Fisica a-Nuclei Particles and Fields 30~(4), 632--652.

\bibitem[{Nayak et~al.(2002)Nayak, Lahiri, and Ramaswami}]{47}
Nayak, D., Lahiri, S., Ramaswami, A., 2002. Alternative radiochemical heavy ion
  activation methods for the production and separation of thallium
  radionuclides. Applied Radiation and Isotopes 57~(4), 483--489.

\bibitem[{Nowotny(1981)}]{48}
Nowotny, R., 1981. Calculation of proton-induced radioisotope production yields
  with a statistical-model based code. International Journal of Applied
  Radiation and Isotopes 32~(2), 73--78.

\bibitem[{NuDat(2011)}]{49}
NuDat, 2011. Nudat 2.5 database http://www.nndc.bnl.gov/nudat2/.

\bibitem[{of-Weights-and Measures(1993)}]{31}
of-Weights-and Measures, I.-B., 1993. Guide to the expression of uncertainty in
  measurement, 1st Edition. International Organization for Standardization,
  Genève, Switzerland.

\bibitem[{Pritychenko and Sonzogni(2003)}]{50}
Pritychenko, B., Sonzogni, A., 2003. Q-value calculator.

\bibitem[{Qaim(2001)}]{51}
Qaim, S.~M., 2001. Nuclear data relevant to the production and application of
  diagnostic radionuclides. Radiochimica Acta 89~(4-5), 223--232.

\bibitem[{Qaim et~al.(1979)Qaim, Weinreich, and Ollig}]{52}
Qaim, S.~M., Weinreich, R., Ollig, H., 1979. Production of 201tl and 203pb via
  proton-induced nuclear-reactions on natural thallium. International Journal
  of Applied Radiation and Isotopes 30~(2), 85--95.

\bibitem[{Rurarz(1994)}]{53}
Rurarz, E., 1994. Excitation functions and yields of proton induced reactions
  at intermediate energies leading to important diagnostics radioisotopes of
  52fe, 77br, 82rb, 97ru, 111in, 123i, 127xe, 128cs, 178ta and 201tl. Tech.
  rep., Soltan Institute for Nuclear Studies, SINS.

\bibitem[{Sakai et~al.(1965)Sakai, Ikegama, Yamazaki, and Saito}]{54}
Sakai, M., Ikegama, H., Yamazaki, T., Saito, K., 1965. Nuclear structure of
  hg200. Nuclear Physics 65, 177--202.

\bibitem[{Sattari et~al.(2003)Sattari, Aslani, Dehghan, Shirazi, Shafie,
  Shadanpour, and Winkel}]{55}
Sattari, I., Aslani, G., Dehghan, M.~K., Shirazi, B., Shafie, M., Shadanpour,
  I.~N., Winkel, P., 2003. Dependence of quality of thallium-201 on irradiation
  data. Iran. J. Radiat. Res. 1, 51--54.

\bibitem[{Sheu et~al.(2003)Sheu, Jiang, and Duh}]{56}
Sheu, R.~J., Jiang, S.~H., Duh, T.~S., 2003. Evaluation of thallium-201
  production in iner's compact cyclotron based on excitation functions.
  Radiation Physics and Chemistry 68~(5), 681--688.

\bibitem[{Shubin(2001)}]{57}
Shubin, Y.~N., 2001. Model calculations and evaluation of nuclear data for
  medical radioisotope production. Radiochimica Acta 89~(4-5), 317--324.

\bibitem[{Shubin et~al.(1998)Shubin, Lunev, Konobeyev, and Dityuk}]{58}
Shubin, Y.~N., Lunev, V.~P., Konobeyev, A.~Y., Dityuk, A.~I., 1998. Mendl-2p
  protonreaction data library for nuclear activation (medium energy nuclear
  data library). Tech. rep., IAEA.

\bibitem[{Szelecs\'enyi et~al.(1995)Szelecs\'enyi, Boothe, Tavano, Fenyvesi,
  and T\'ark\'anyi}]{59}
Szelecs\'enyi, F., Boothe, T.~E., Tavano, E., Fenyvesi, A., T\'ark\'anyi, F.,
  1995. Evaluation of cross sections / thick target yields for 201tl
  production. In: Link, J.~M., R.-T.~J. (Ed.), Sixth Workshop on Targetry and
  Target Chemistry. TRIUMF, Publishing Office, p. 127.

\bibitem[{Tak\'acs et~al.(2005)Tak\'acs, T\'ark\'anyi, and Hermanne}]{60}
Tak\'acs, S., T\'ark\'anyi, F., Hermanne, A., 2005. Validation and upgrading of
  the recommended cross-section data of charged particle reactions: Gamma
  emitter radioisotopes. Nuclear Instruments \& Methods in Physics Research
  Section B-Beam Interactions with Materials and Atoms 240~(4), 790--802.

\bibitem[{T\'ark\'anyi et~al.(2001)T\'ark\'anyi, Tak\'acs, Gul, Hermanne,
  Mustafa, Nortier, Oblozinsky, Qaim, Scholten, Shubin, and Youxiang}]{61}
T\'ark\'anyi, F., Tak\'acs, S., Gul, K., Hermanne, A., Mustafa, M.~G., Nortier,
  M., Oblozinsky, P., Qaim, S.~M., Scholten, B., Shubin, Y.~N., Youxiang, Z.,
  2001. Beam monitor reactions (chapter 4). charged particle cross-section
  database for medical radioisotope production: diagnostic radioisotopes and
  monitor reactions. Tech. rep., IAEA.

\bibitem[{Tel et~al.(2011)Tel, Sahan, Aydin, Sahan, Ugur, and Kaplan}]{62}
Tel, E., Sahan, M., Aydin, A., Sahan, H., Ugur, F.~A., Kaplan, A., 2011. The
  Newly Calculations of Production Cross Sections for Some Positron Emitting
  and Single Photon Emitting Radioisotopes in Proton Cyclotrons. InTech,
  http://www.intechopen.com/books/radioisotopes-applications-in-physical-sciences/the-newly-calculations-of-production-cross-sections-for-some-positron-emitting-and-single-photon-emi.

\bibitem[{Wauters et~al.(1987)Wauters, Vandecasteele, Strijckmans, and
  Hoste}]{63}
Wauters, G., Vandecasteele, C., Strijckmans, K., Hoste, J., 1987. Determination
  of cadmium, thallium and lead in environmental-samples by proton
  activation-analysis. Journal of Radioanalytical and Nuclear
  Chemistry-Articles 112~(1), 23--31.

\bibitem[{Zaitseva et~al.(1987)Zaitseva, Kovalev, Knotek, Khalkin, Ageev,
  Klyuchnikov, and Linev}]{64}
Zaitseva, N.~G., Kovalev, A.~S., Knotek, O., Khalkin, V.~A., Ageev, V.~A.,
  Klyuchnikov, A.~A., Linev, A.~F., 1987. Preparing 201tl from lead bombarded
  by protons of energy about 65 mev. Soviet Radiochemistry 29~(2), 235--240.

\end{thebibliography}







\end{document}